\documentclass[12pt]{aastex} 

\usepackage{emulateapj5}

\newcommand{\ra}[4]{${\rm RA}={#1}^{\rm h}{#2}^{\rm m}{#3}\fs {#4}$}
\newcommand{\dec}[4]{${\rm Dec.}={#1}^{\circ}{#2}\arcmin{#3}\farcs
{#4}$}

\newcommand{\e}[1]{$10^{#1}$} 
\newcommand{\cm}[1]{\,cm$^{#1}$}
\newcommand{\ee}[1]{$\times 10^{#1}$}

\newcommand{\erg}{~erg\,cm$^{-2}$\,s$^{-1}$}

\newcommand{\mjb}{\,mJy\,beam$^{-1}$} 
\newcommand{\hr}{H\,\textsc{ii}}

\begin{document}

\title{A high-resolution study of nonthermal radio and X-ray  emission
 from SNR G347.3--0.5}

\author{J. S. Lazendic\altaffilmark{1}, P. O. Slane\altaffilmark{1},
B. M. Gaensler\altaffilmark{1}, S. P. Reynolds\altaffilmark{1,2}, \\ 
P. P. Plucinsky\altaffilmark{1} and J. P. Hughes\altaffilmark{3}} 

\altaffiltext{1}{Harvard-Smithsonian Center for Astrophysics, 60
Garden Street, Cambridge MA 02138}
\altaffiltext{2}{Physics Department, North Carolina State University, 
Raleigh, NC 27695-8202}
\altaffiltext{3}{Department of Physics and Astronomy, Rutgers University, 136
Frelinghuysen Road, Piscataway, NJ 08854}

\vfil

\begin{abstract}

 G347.3--0.5 is one of three shell-type supernova remnants in the
 Galaxy whose X-ray spectrum is dominated by nonthermal emission.
 This puts G347.3--0.5 in the small, but growing  class of SNRs for
 which the X-ray emission reveals directly the presence of extremely
 energetic electrons accelerated by the SNR shock.  We have obtained
 new high-resolution X-ray and radio data on G347.3--0.5 using the
 {\em Chandra X-ray Observatory} and the Australia Telescope Compact
 Array (ATCA) respectively.  The bright northwestern   peak of the SNR
 seen in {\em ROSAT} and {\em ASCA} images is resolved with {\em
 Chandra}  into bright filaments and fainter diffuse emission.  These
 features show good correspondence with the radio morphological
 structure, providing strong evidence that the same population of
 electrons is responsible for the synchrotron emission in both bands
 in this part of the remnant.  
 Spectral index information from both observations is presented.  
 We found significant difference in photon index value between bright and
 faint regions of the SNR shell. Spectral properties of these regions 
 support the notion that efficient particle acceleration is 
 occurring in the bright SNR filaments.  We report the detection 
 of linear radio polarization towards the SNR, which is most ordered
 at the northwestern  shell where particle acceleration is presumably
 occurring. Using our new {\em Chandra} and ATCA data we model the 
 broad-band emission from G347.3--0.5 with the synchrotron 
 and inverse Compton mechanisms and discuss the conditions under 
 which this is a plausible scenario.

\end{abstract}

\keywords{acceleration of particles ---  cosmic rays --- radiation
 mechanisms: nonthermal --- supernova remnants: individual:
 G347.3--0.5 --- X-rays: general }

\section{INTRODUCTION}

 Due to the release of an enormous amount of energy ($\sim 10^{51}$
 erg)  at their creation,   supernova remnants (SNRs) have long been
 considered as a primary source of  Galactic cosmic rays with energies
 up to  the ``knee'' of the spectrum at $\sim 3 \times 10^{15}$~eV
 \citep{shklovsky53,ginzburg57}. Cosmic rays  with energies higher
 than this are believed to be extragalactic in origin
 \citep{axford94}.   First-order Fermi shock acceleration, also called
 diffusive shock acceleration, in which particles gain energy from
 scattering back and forth across the shock, has been suggested as the
 most probable acceleration  mechanism in SNR shocks
 \citep[see][]{reynolds81,blanford87,jones91}. However, until recently
 the observational evidence for the production of high  energy
 particles in SNRs was poor and came mainly from the fact that SNRs
 emit synchrotron radiation in the radio band.   The major
 observational break-through came only recently with  the detection of
 nonthermal X-ray emission from the shell-type SNR SN~1006
 \citep{koyama95}. The featureless X-ray spectrum found  at the rim of
 SN~1006, in contrast to the thermal spectra found towards the
 interior of the remnant, was fitted well with a power law of photon
 index $\Gamma \sim 2.2$ (where the photon flux $F$ obeys 
 $F \propto (h\nu)^{-\Gamma}$).  Further evidence came 
 with the detection
 of TeV $\gamma$-ray emission from  SN~1006 \citep{tanimori98}.  There
 are several mechanisms capable of producing TeV energy photons,
 including inverse Compton (IC) scattering, nonthermal bremsstrahlung
 and pion-decay.  Broad-band modeling of the emission from SN~1006
 indicates that IC scattering is responsible for the TeV $\gamma$-ray
 emission \citep[e.g.,][]{mastichiadis96,allen01}. However, cosmic rays
 are comprised mostly of protons and there is not yet clear evidence
 of proton acceleration in SNRs, aside from the suggestion that the
 TeV emission results  from neutral pion-decay
 \citep{aharonian99,enomoto02}.  Another problem in the quest for a
 cosmic ray  origin in SNRs is that the maximum energy of electrons
 produced by SNRs seems to  fall below the ``knee''. For example,
 studies of about 20  SNRs with mostly thermal X-ray emission imply
 that the maximum energy to which electrons can be accelerated does
 not exceed \e{14} eV \citep{reynolds99,hendrick01}. 

Two more shell-type remnants with dominant non-thermal X-ray spectra
have been identified:  G347.3--0.5 \citep{koyama97,slane99} and
G266.2--1.2 \citep{slane01}. G347.3--0.5 (RX~J1713.7--3946) 
was first discovered in the  {\em ROSAT
(R\"{o}ntgensatellit)}  All-Sky Survey by \citet{pfeffermann96}, who
used a thermal plasma  model to infer a very high temperature of $kT \sim
4.8$ keV, and column density of $N_H \sim 4.5 \times 10^{21}$ \cm{-2}.
Based on this column density the  distance to the SNR  was estimated
to be $\sim 1.1$ kpc, while the derived plasma temperature implied an
SNR age of $\sim 2100$ yrs.   However,  subsequent {\em ASCA (Advanced
Satellite for Cosmology and Astrophysics)} observations revealed that
the X-ray emission from the remnant is predominantly nonthermal
\citep{koyama97,slane99}.  The remnant is $\sim 1$\degr\ in diameter
and appears to be of a shell--type morphology with the brightest
emission in the western region.  The SNR is located on the edge of the
molecular cloud complex that encompasses the \hr\ region G347.61+0.20
located northwest of the SNR (see Figure~\ref{fig-atca}). 
 Assuming that the SNR is physically
associated with the molecular cloud complex,  the  distance was
estimated to be  $6.3 \pm 0.4$~kpc from existing observations of the
CO 1--0  line emission towards the complex \citep{slane99}. {\em ASCA}
observations did not reveal any line emission from the SNR interior,
and this lack of thermal emission sets an upper limit on  the mean
density around the remnant of $< 0.3$ \cm{-3} \citep{slane99}.
Such a low density suggests that the majority of the remnant is still
evolving in the interior of the large circumstellar cavity driven by
the wind of its massive progenitor.  In the northwestern region of
G347.3--0.5, where the SNR may be interacting with denser molecular
gas, the upper limit on the ambient density is higher ($< 1$
\cm{-3}),  which is broadly consistent with the densities estimated
around other SNRs that  are associated with molecular clouds.  
Most recently, \citet{pannuti03} reported detection of
 a thermal component towards the center of the SNR, which implies
gas density of 0.05--0.07\cm{-3} in this part of the SNR.

Two point sources have been identified within the boundaries of  the
remnant shell:  1WGA~J1714.4--3945 is believed to be of stellar origin
\citep{pfeffermann96}, while 1WGA~J1713.4--3949 is located at the
center of the SNR, has no obvious optical counterpart, and is  a
candidate for an associated neutron star \citep{slane99}.  Our {\em
Chandra} observations also included the latter point source, results
for which are presented elsewhere \citep{lazendic03}.

G347.3--0.5 was also detected at TeV energies  with the
CANGAROO\footnote{Collaboration of Australia and Nippon for a
Gamma-ray Observatory in the Outback} telescope
\citep{muraishi00,enomoto02}, and models of broad-band  emission point
to IC scattering as the origin of the  TeV photons
\citep{muraishi00,ellison01}.  More recently, follow up TeV
$\gamma$-ray  observations have led to a different conclusion,
suggesting pion-decay as the source of energetic photons
\citep{enomoto02},  but the nature of this emission is still
uncertain \citep[see][]{butt02,reimer02}.  An EGRET\footnote{Energetic
Gamma Ray Experiment Telescope}  source 3EG~J1714--3857
\citep{hartman99}  was also detected  near the remnant and linked to
the SNR interaction with a molecular cloud \citep{butt01}.  This is
potentially  supported by the identification of the hard X-ray source
AX~J1714.1--3912, which appears to coincide with the location of  one
of the  molecular clouds in the complex, and possibly with the EGRET
source \citep{uchiyama02}. However,  the energetics required to yield
the observed X-ray flux of the source appear problematic relative to
the total energy budget of the SNR unless the source distance is a
factor of five smaller than suggested by the molecular line velocity
of the cloud. 

To investigate the fine-scale structure of the SNR and
its relationship with particle acceleration,
we obtained high-resolution X-ray and
radio data on G347.3--0.5 using the {\em Chandra  X-ray Observatory}
and the Australia Telescope Compact Array (ATCA).  
Our goal was to look for
spectral variations in different spatial regions, to improve the
estimates on thermal emission, to compare the X-ray
morphology with high-resolution radio images, to correlate the
radio and X-ray spectral indices and to investigate linear
polararization.   We present these results here,
along with modeling of the broad-band spectrum of G347.3--0.5 to
investigate the  origin of the accelerated high energy particles in
this SNR. 

In section $\S$ 2 we describe our X-ray observations,
 and present images and spectral results for the SNR.  Radio
 observations and images  of the SNR are presented in section $\S$3,
 where we also  derive a radio spectral index and measure linear
 polarization in the SNR. In section $\S$4 we discuss  
the spectral variations across the nortwestern SNR shell and in
 section $\S$5 we investigate the SNR
 morphology and the correlation between X-ray and radio images.  In
 section $\S 6$ we present synchrotron and inverse Compton modeling of the
 broad-band spectrum from G347.3--0.5  and discuss the origin of the
 TeV emission in this SNR.

\section{X-RAY DATA}

G347.3--0.5 was observed with the Advanced CCD Imaging Spectrometer
(ACIS) detector on board the {\em Chandra X-ray Observatory} on 2000
July 25.  ACIS consists of two CCD arrays --- the ACIS-I (Imaging)
array has four CCDs arranged in a square, while the ACIS-S
(Spectroscopic) array has six linearly adjacent CCDs. Only six CCDs
can be used at one time.  The angular size of G347.3--0.5 is much
larger than the ACIS-I field of view  ($16\farcm 9 \times 16\farcm
9$),  so we observed two diametrically opposed fields covering regions
of the SNR shell, as shown in  Figure~\ref{fig-chandra+rosat}.  Field
1 (ObsId 736) was centered on the bright northwestern rim at
\ra{17}{11}{49}{8}, \dec{-39}{36}{18}{0}\ and consisted of a 30 ks
exposure. Field~2  (ObsId 737) was observed for 40 ks, positioned at
the fainter eastern SNR rim centered at \ra{17}{15}{16}{5},
\dec{-39}{57}{56}{0}. Two ACIS-S chips were also turned on: for field
1, the ACIS-S1 and ACIS-S3 chips were used to include the central
point source and a region in the SNR interior,  while for field~2
ACIS-S2 and ACIS-S3 were used to cover the eastern SNR rim. Data were
taken in full-frame timed--exposure (TE) mode with the standard
integration time  of 3.2 s.

Data were reduced using standard threads in the  {\em Chandra} {\tt
Interactive Analysis of Observations (CIAO)} software package version
2.2.1.  To mitigate the degradation in the spectral response of the
ACIS chips caused by radiation damage early in the mission
\citep[see][]{prigozhin00}, we also used software
\citep{townsley00,townsley02} to correct for the effect of increased
charge transfer inefficiency (CTI), such as gain variations, event
grade distortion and degraded energy resolution as a function of row
number in each CCD. These corrections increase   the number of
detected events and improve the assigned event energies and   spectral
resolution.  We applied CTI corrections to the  Level 1 processed
event list provided by the pipeline processing done  at the {\em
Chandra} X-ray Center (CXC). The data were then screened  for
``flaring'' pixels and filtered with standard {\em ASCA} grades
(02346). The effective exposure time after data processing  was 29.6
ks and 38.9 ks for field~1 and 2, respectively. 

\subsection{X-ray Image}

\begin{figure}
\epsscale{1.0}
\plotone{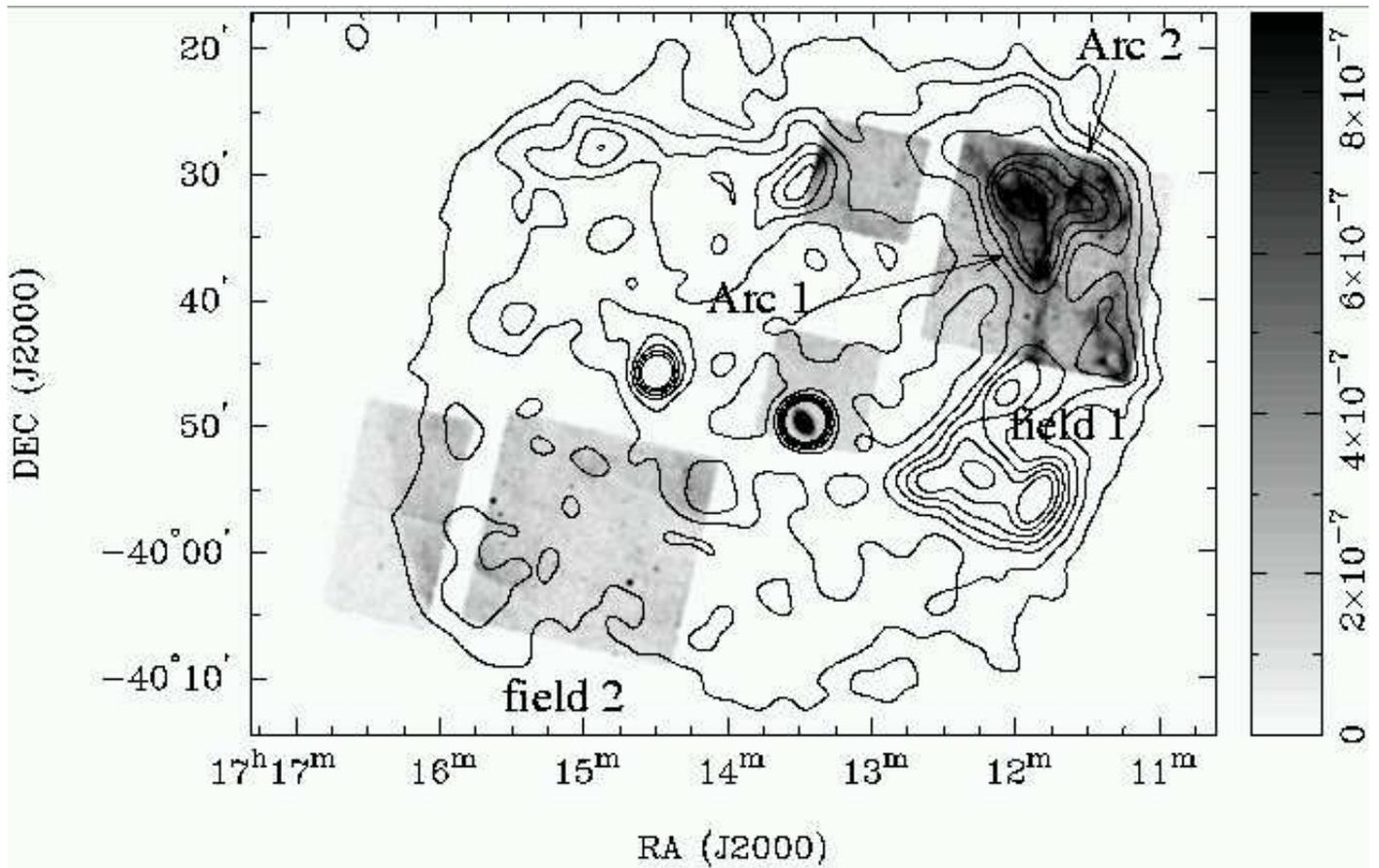}
\caption{ The 1.0--8.0 keV band {\em Chandra} ACIS greyscale image of
G347.3--0.5, including the central source 1WGA~J1713.4--3949 which
appears elongated due to off-axis location in the ACIS-S1 chip 
\citep[see][for details on this source]{lazendic03}.  The
linear greyscale is in units counts\,arcsec$^{-2}$\,s$^{-1}$ and is
saturated to show regions with low emission.  To show the whole extent
of the remnant, the {\em Chandra}  image is overlaid with the  {\em
ROSAT} PSPC contours \citep{slane99}.  Contour levels are: 7, 12, 17,
22, 27, 32, 37 $\times$ 0.018 counts\,arcmin$^{-2}$\,s$^{-1}$.  Two
bright arcs, observed in radio band (see Figure~\ref{fig-atca}),  are
labeled: Arc~1, of which {\em Chandra} covered southwestern portion,
extends approximately north-south,  while Arc~2  is almost
perpendicular to Arc~1. The two observed  {\em Chandra} fields are
also labeled as field~1 and field~2.}
\label{fig-chandra+rosat}
\end{figure}

We used the {\tt CIAO} script {\tt merge\_all} to combine the data from the
two  fields into an exposure-corrected X-ray image in the energy range
1--8 keV. Since the effective area of the detector depends both on the
energy of incident photons and their positions  on the CCD, we applied
spectral weighting to the merged image, which takes into account the
fraction of the incident flux  falling in the particular part of the
band. The exposure-corrected image was blanked where the exposure
was less than 15\% of its maximum value. The final image is shown in
Figure~\ref{fig-chandra+rosat} overlaid with the {\em ROSAT}
contours. The image was binned in $4\farcs 6 \times 4\farcs 6$ pixels
and smoothed with a Gaussian filter with a FWHM of 2\arcsec.  
Since CTI corrections were not available for ACIS-S1 chip on which 
the central source is located, these data were  processed separately
and then added to the image. We also made soft
(0.5--2.1 keV) and hard band (2.1--8.0 keV) images, but they do not
show any significant brightness  variation with energy. 

The northwestern peak seen in {\em ROSAT} and {\em ASCA} images is
resolved with {\em Chandra} observations into two bright arcs and
fainter diffuse emission, as shown in  Figure~\ref{fig-chandra+rosat}.
As we discuss below, this structure bears significant resemblances to
what is  seen in radio images.  Arc~1 (see
Figure~\ref{fig-chandra+rosat} and Figure~\ref{fig-regions})  appears to
be made of thin filaments delineating the SNR shock front, while Arc~2
has  a more irregular shape. Diffuse emission along the eastern shell
does not show significantly different structure from that seen with
{\em ROSAT}, although the limb is more clearly defined and extends
beyond contours obtained from the {\em ROSAT} image.

\subsection{X-ray Spectra}
\label{sec-xray}

\begin{figure}
\epsscale{0.7}
\plotone{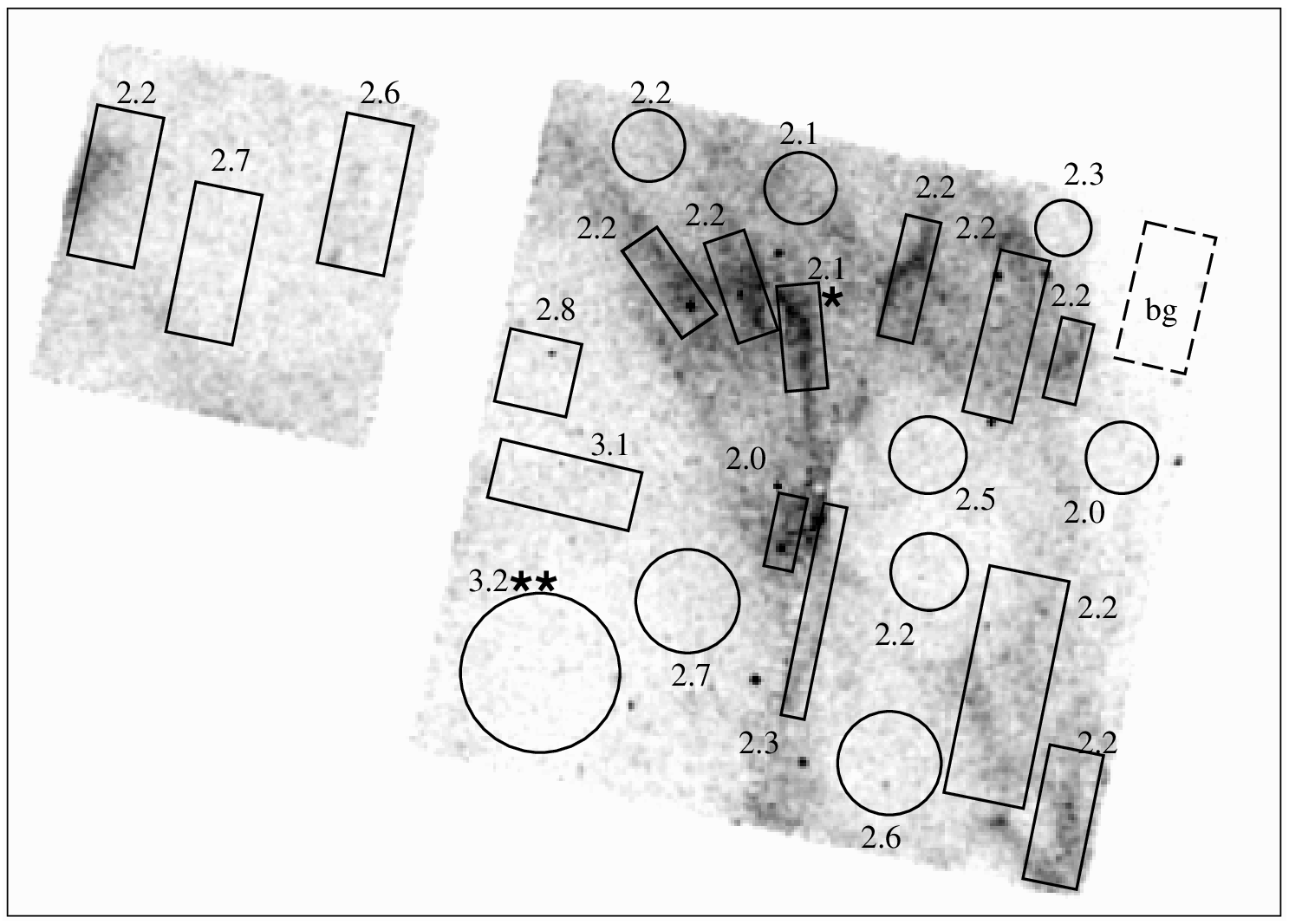}
\plotone{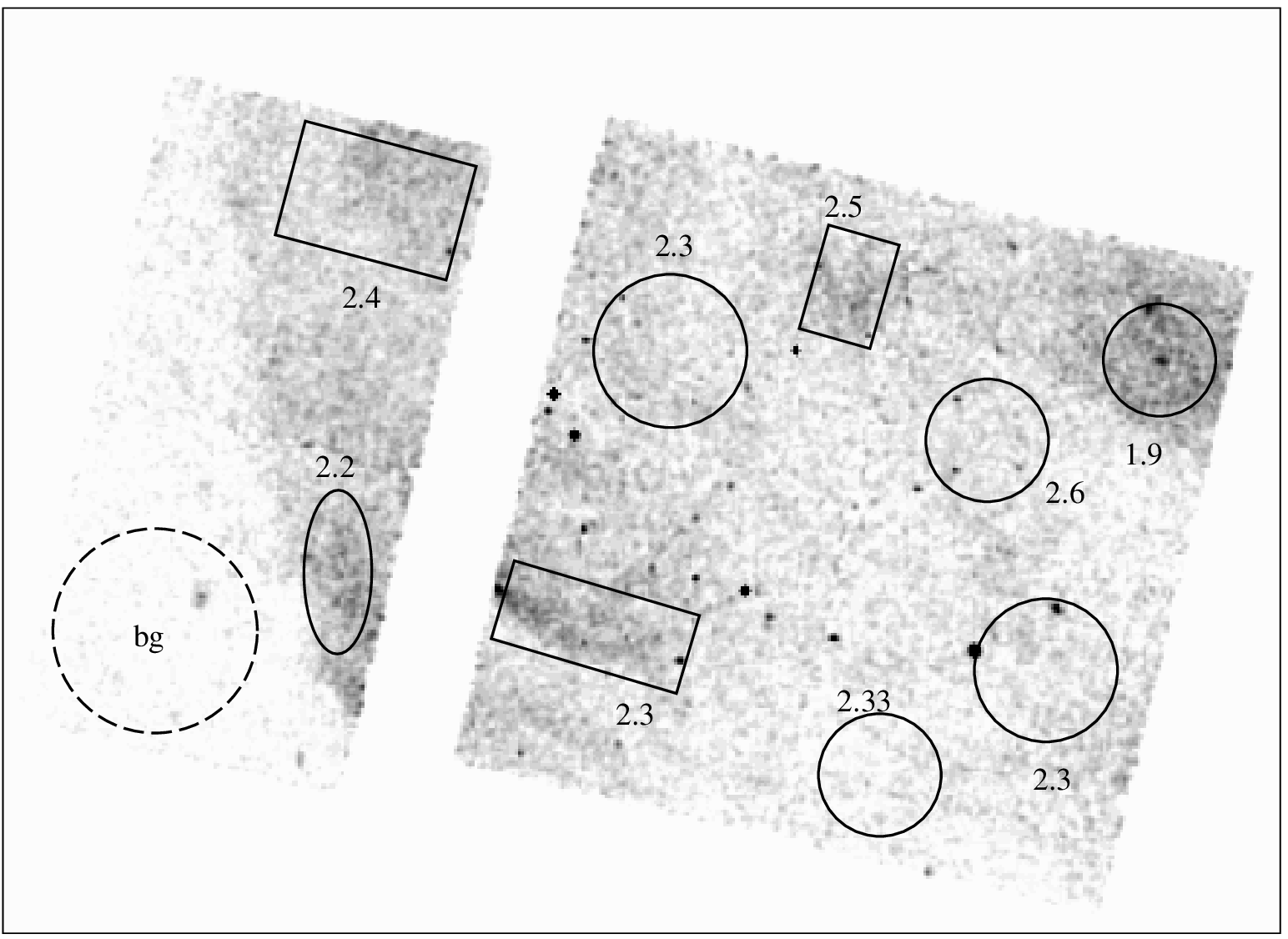}
\caption{Extraction regions from field~1 and 2 and their corresponding
 best-fit photon index values. The regions outlined
with the dashed line are used for background subtraction.}
\label{fig-regions}
\end{figure}

X-ray spectra were extracted from discrete regions of the  ACIS chips
in field~1 and  field~2, shown in Figure~\ref{fig-regions}.   For each
spectrum the counts were  grouped with a minimum of 25 counts per
spectral channel.  We use redistribution matrix files (RMF)
appropriate for CTI-corrected data \citep{townsley00,townsley02},  and
created weighted auxiliary response files (ARF) applying the same
spectral binning as used in the corresponding RMFs.   The background
for the spectral fitting was extracted from the northwest corner in
field~1 and from the southeast corner in field~2  (see 
Figure~\ref{fig-regions}).

Shock acceleration of electrons to X-ray emitting energies is expected
to produce a roughly power law distribution with an exponential cutoff
above some maximum energy, resulting in a photon spectrum rolling
slowly off through the X-ray band. The {\em Chandra} bandpass is not large
enough in most cases to distinguish this slight curvature from a
strait power law. The X-ray slope bears no particular relation to
 the slope at lower frequencies, but can serve as an indication of the
location of the rolloff. We shall use both power law fits (giving
photon index $\Gamma$) and cutoff-model fits ({\tt SRCUT}, giving a
rolloff frequency) to describe our spectra. 

\begin{figure}
\epsscale{0.6}
\plotone{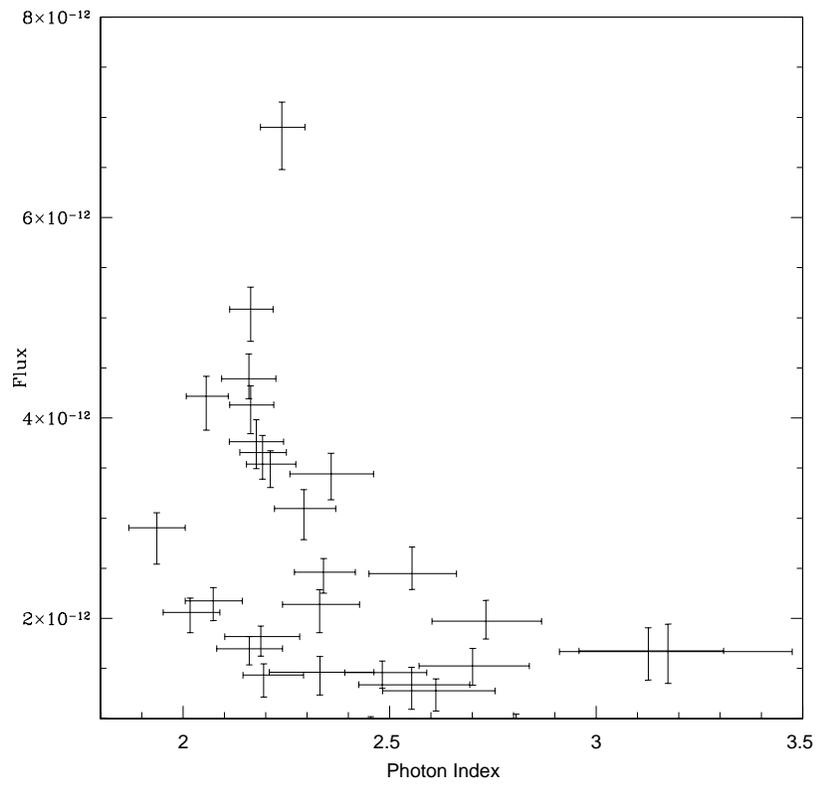}
\plotone{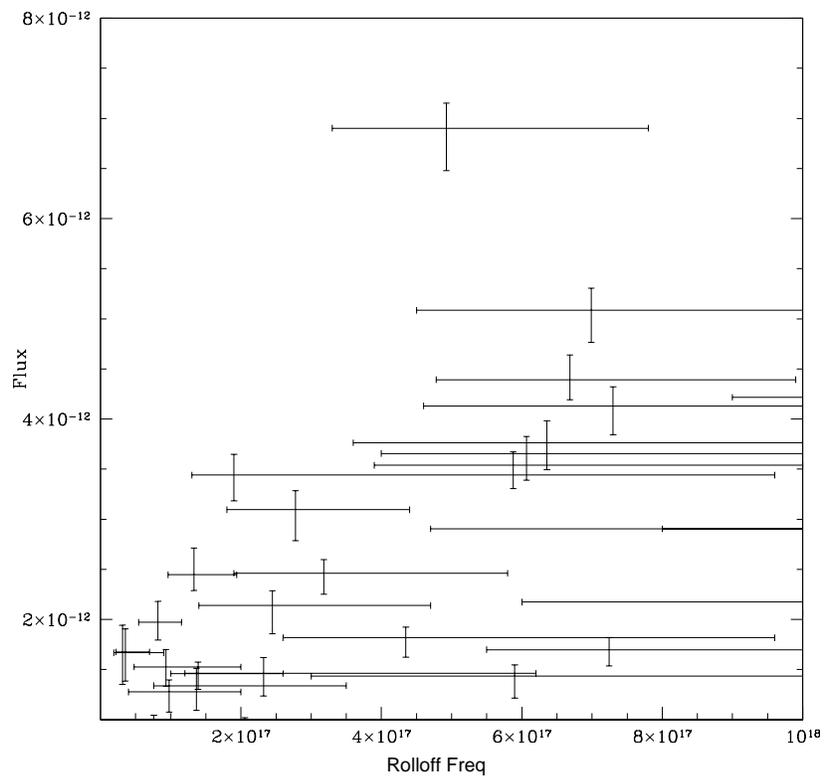}
\caption{The best-fit values of X-ray flux and  photon index/rolloff
 frequency   values  with 90\% confidence level derived from power law 
 and {\tt SRCUT} model fits to the spectra from regions 
 in field~1 and field~2 marked in Figure~\ref{fig-regions}. }
\label{fig-fg}
\end{figure}

\begin{figure}
\epsscale{1.0}
\plotone{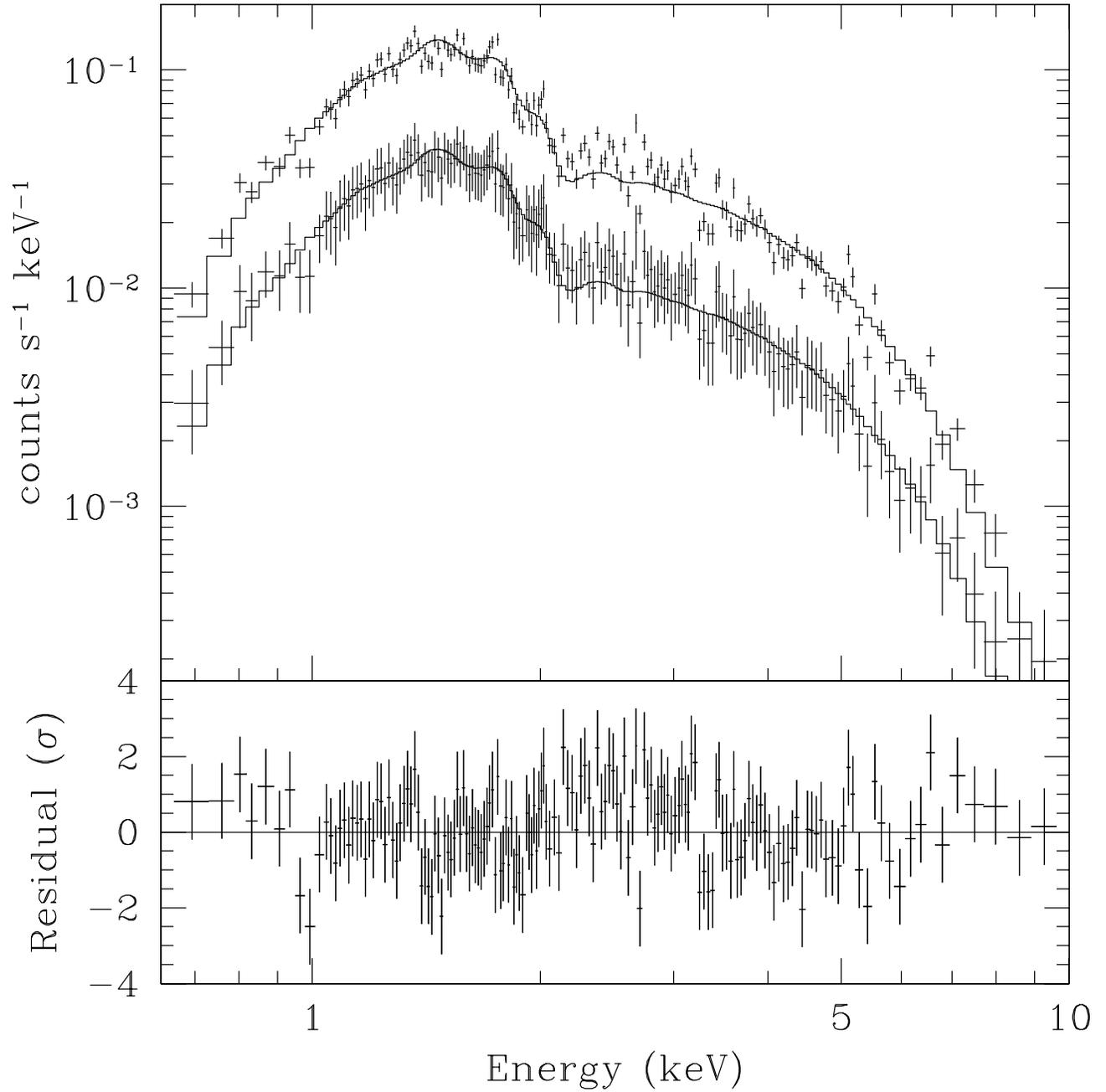}\\
\caption{Representative ACIS spectra from {\em Chandra} observations
of  G347.3--0.5 and the residuals for the best-fit model. The top
spectrum is extracted from one of the brightes regions in SNR, 
 marked with a star in Figure~\ref{fig-regions}, and the bottom spectrum
 is from one of the faint regions marked with two stars in
Figure~\ref{fig-regions}.}
\label{fig-spectra}
\end{figure}

Individual  spectra were first fitted separately to search  for any
change in the  photon index across the source and  for the presence of
thermal component. We found small-scale spectral differences in
different regions.  We could fit those either with variations in
column density, and roughly the same photon index ($\sim 2.1$), or, if
the column density were fixed, with variations in the photon index.
 In the first case, we found that the column density was higher 
 towards the bright regions ($\sim 8\times10^{21}$\cm{-2}) than
towards the fainter regions ($\sim 5\times10^{21}$\cm{-2}). Thus, the
brighter regions have higher column density, which is opposite
 from what one expects if the brightness variation is caused
by varying interstellar absorption. Furthermore, the fact that the 
radio and X-ray emission have the similar morphologies in field~1 
implies that X-ray brightness distribution is not the result of
variations in absorption. Thus, it seems plausible that 
 the column density is constant across field~1 and  we adopt the second 
 case, fixing the column density at the value found
in the spatially averaged fitting and allowing $\Gamma$ to vary. 
Photon index values derived with the second approach 
are plotted in Figure~\ref{fig-fg} and  imply that
there is significant spectral variation between  the bright and the
faint SNR regions. The bright emission  has values $2.05 < \Gamma <
2.25$, while the faint emission  takes broad range of values, 
with the steepest photon index of $\sim 3$.  Representative spectra
with power law model fit are shown in Figure~\ref{fig-spectra}. The average 
values of photon index across field~1 of $\sim 2.35$
is consistent with that of $2.41$ for the northwestern region  from
the previous study with {\em ASCA} \citep{slane99}.

We then fitted our data with the {\tt SRCUT} model
\citep{reynolds99}, which calculates a synchrotron spectrum 
from a power law distribution of electrons modified by an exponentially cut
off in a uniform magnetic field. The model can be used to 
derive the maximum rolloff frequency $\nu_{rolloff}$ related 
to the maximum electron energy. Derived $\nu_{rolloff}$ values are
plotted  in Figure~\ref{fig-fg}. We frozen the radio spectral index $\alpha$
(where the radio flux $S_{\nu} \propto \nu^{-\alpha}$) parameter
for each spectrum to the global SNR value of 0.6 
and left normalization (i.e., the 1~GHz  radio flux density) as a free
parameter. The model gave reduced $\chi^2$ value comparable to that of
the power law model and the range of values for rolloff frequency
between 3\ee{16}--2\ee{18}~Hz, with the fainter regions having, in
generall, lower  $\nu_{rolloff}$ values.

The SNR spectra from regions we observed with {\em Chandra} show no
evidence for any emission lines. Adding equilibrium or non-equilibrium
thermal models to the power law fits of the individual or joint fit to
the spectra does not improve the fit significantly.  
However, spectra from some regions in both ACIS-S3 chips show 
a possible line feature around 0.8~keV. Modeling this feature does not
provide strong constraints on the thermal emission properties, so 
 deeper observations are needed to follow up this
potential detection of thermal emission in G347.3--0.5.

\section{RADIO DATA}

Radio observations of G347.3--0.5 were obtained with the ATCA during
January, March and April 1998. The array consists of six  22~m
antennas that can be configured to give baselines between 31~m and 
6~km \citep[see][]{frater92}.  We used the array in three different
configurations (375, 750 A and 1.5 A) to provide optimal 
$uv$-coverage of the observed region. Due to the large extent of the SNR
and the  complexity of the field around it, the region was imaged in
mosaic mode with 10 pointing centers.  Data were taken simultaneously
at two frequencies, 1.4 and 2.5~GHz, each using a bandwidth of 128~MHz
split in 32 4-MHz channels.   For all the observations the primary
calibrator PKS~B1934--638 was used for  bandpass and absolute flux
calibration. PKS~1740--517 was used as a  secondary calibrator for
antenna gains and instrumental polarization calibration. 

\subsection{Radio Images}

Radio data were reduced using standard procedures of the {\tt MIRIAD}
software package \citep{sault99}. In the editing procedure, 4 channels
on each side of the band were discarded and the remaining  26 central
channels were  averaged down to 13 8--MHz channels by Hanning
smoothing. A uniform weighting was applied
to visibilities  to minimize sidelobes in the images.   The longest
baselines (i.e., all correlations with the sixth antenna) were excluded
to enhance the surface brightness sensitivity.  Also, amplitude and
phase self-calibration were applied to the 1.4~GHz data to  improve
solutions for antenna gains using a strong point source in the eastern
edge of the observed field  (outside the image shown here).  The
images were made using  the multi-frequency synthesis \citep{sault94}
and deconvolved using the mosaiced maximum entropy approach
\citep{sault96}.  The resulting images were convolved with a Gaussian
restoring beam listed in Table~\ref{tab-atca}, and
corrected for primary beam response. The rms sensitivity in the images
is also listed in Table~\ref{tab-atca}.

\begin{figure}
\epsscale{1.0}
\plotone{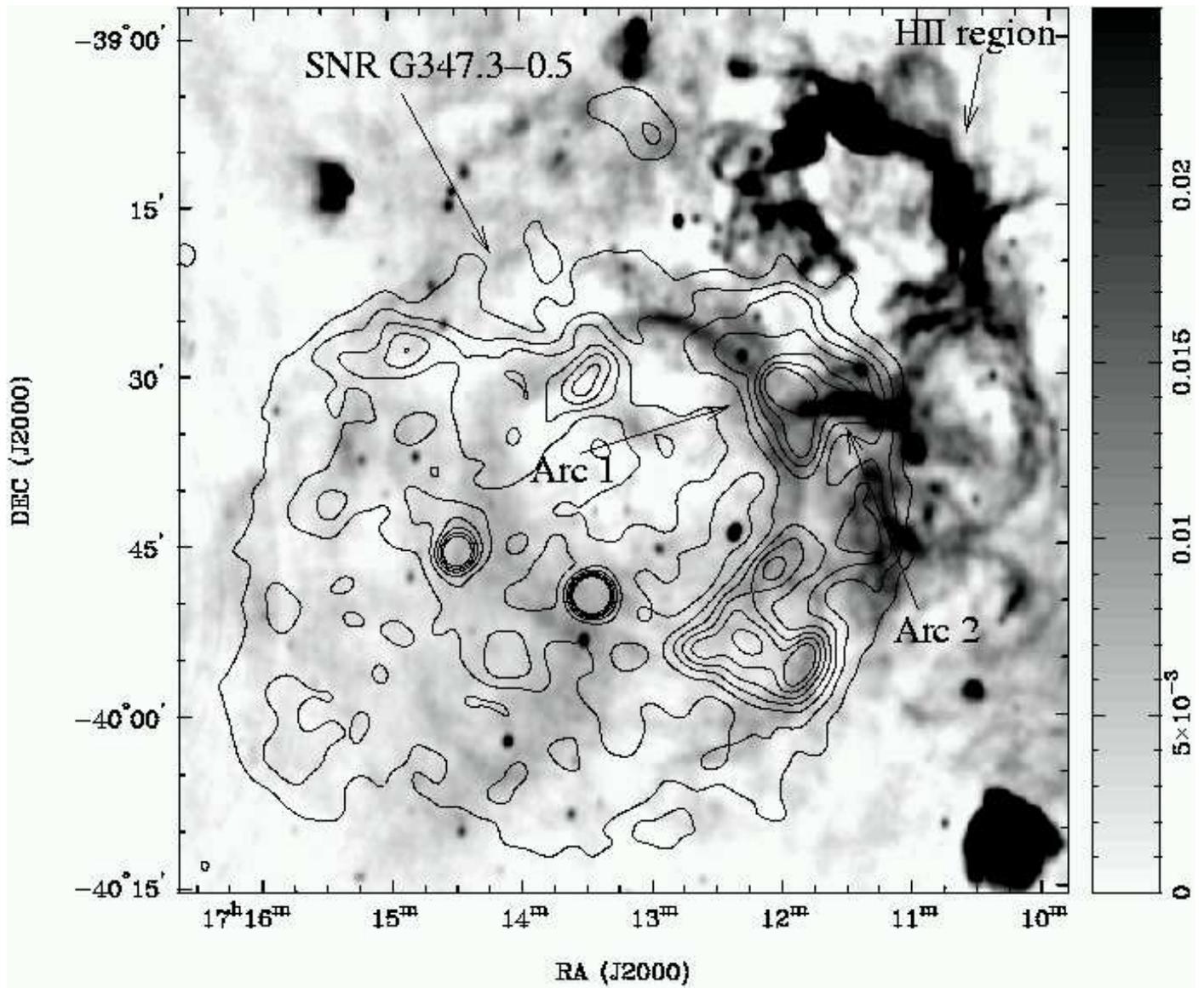}
\caption{ATCA images of G347.3--0.5 and surrounding region  at 1.4~GHz. The image was convolved with a Gaussian  restoring beam of
$46\arcsec \times 36\arcsec$  (P.A.= $-$3.8), shown by the tiny
ellipse in the bottom left corner.  The image is overlaid with the
{\em ROSAT} contours with the levels same as in Figure~\ref{fig-chandra+rosat}.
The linear grayscale is in units Jy\,beam$^{-1}$.}
\label{fig-atca}
\end{figure}

In Figure~\ref{fig-atca} we show the image at 1.4~GHz overlaid with
the {\em ROSAT} contours.  As noted by
\citet{slane99}, in the radio band the SNR appears as a faint shell
with two bright arcs to the west (marked in
Figure~\ref{fig-atca}). A faint inner ring of emission is
also evident in the radio image of the SNR (Figure~\ref{fig-atca}),  
with a diameter of $\sim 30\arcmin$.  The
image at 1.4~GHz has angular resolution comparable to the Molonglo
Observatory Synthesis Telescope (MOST)  image \citep{slane99} and was
published in preliminary form by  \citet{ellison01}. While the image
at 2.5~GHz has improved angular resolution, it shows no significant
morphological  difference with respect to the other two images, and
thus we do not show it here. 

We find no obvious radio source at the location of the  {\em ASCA}
source AX~J1714.1--3912, believed to be associated with the EGRET
source 3EG~J1714--3857 in the vicinity of the SNR \citep{uchiyama02}.
There are, however, a few compact radio sources around that location
which are probably thermal \hr\ regions.

\subsection{Radio Spectral Index}

To determine the spectral index of the bright western SNR rim, we used
the ``spectral tomography'' approach \citep{katz97}.  Since source
visibilities at different frequencies are sampled with  different
$u-v$ spacings, we first had to resample the 1.4~GHz data  to match
the $uv$-coverage of the 2.5~GHz data  \citep[see
e.g.,][]{crawford01}.  We first removed the primary beam attenuation 
from the 1.4~GHz image, to which we then applied the primary beam
attenuation and mosaic pattern  of the 2.5~GHz data. The resulting
image was Fourier transformed and resampled using the transfer
function  of the 2.5~GHz data. We then imaged and deconvolved
these modified 1.4~GHz data using the same procedures as for the
2.5~GHz data.  The modified 1.4~GHz and original 2.5~GHz image were
then  both smoothed to a resolution of 60\arcsec. The 2.5~GHz image
was  scaled by a trial spectral index, $\alpha_t$, and subtracted from
the 1.4~GHz image, $I_{\alpha_t}=I_{1.4}-(1.4/2.5)^{\alpha_t}
I_{2.5}$,  where $I_{\alpha_t}$ is the difference image, and $I_{1.4}$
and $I_{2.5}$ are the images at 1.4 and 2.5 GHz, respectively. The
spectral index  $\alpha$ is then found
as the value of $\alpha_t$  at which a particular feature in the image
blends into the background, while the range of  $\alpha_t$  values for
which the residuals of the blended feature  are significant gives the
uncertainty in the spectral index.  

We find a spectral index of $0.50 \pm 0.40$ for the bright northwestern
region of the SNR.  This large uncertainty in
the spectral  index determination is caused by calibration errors and
dynamic range limitations in bright, complex regions.  The spectral
index of Arc~1 is $0.45 \pm 0.30$ and of Arc~2 is $0.10 \pm 0.40$. 
The radio emission from other regions in the SNR is generally
too faint for spectral index determination.  The spectral index of the
 large \hr\ region north-west from the SNR is found to be $-0.10 \pm 0.05$.

\subsection{Radio Polarization}

Continuum observations with the ATCA provide simultaneos 
recording of all four Stokes parameters. We therefore made
images of the $Q$, $U$ and $V$ Stokes parameters at 1.4 
and 2.5~GHz for each of the
13 data channels to minimize bandwidth  depolarization. To produce an
image of polarized intensity we used the {\tt MIRIAD} task
 {\tt pmosmem} which  performs a joint maximum entropy  deconvolution of the
total and polarized intensities for mosaic  observations
\citep{sault99-2}. Cleaned $Q$ and $U$ images  
were then restored with a Gaussian beam, listed in
 Table~\ref{tab-atca}, and corrected for
primary beam response.  A linear polarization image, 
$P_I=(Q^2+U^2)^{1/2}$,  was
formed for each pair of the Stokes $Q$ and $U$ images  and corrected
for non-Gaussian noise statistics \citep{killeen86}.  The mean of the
13 $P_I$ planes was formed and then blanked  
where total or polarized emission fell below the 5$\sigma$ level.

\begin{figure}
\epsscale{1.0}
\plotone{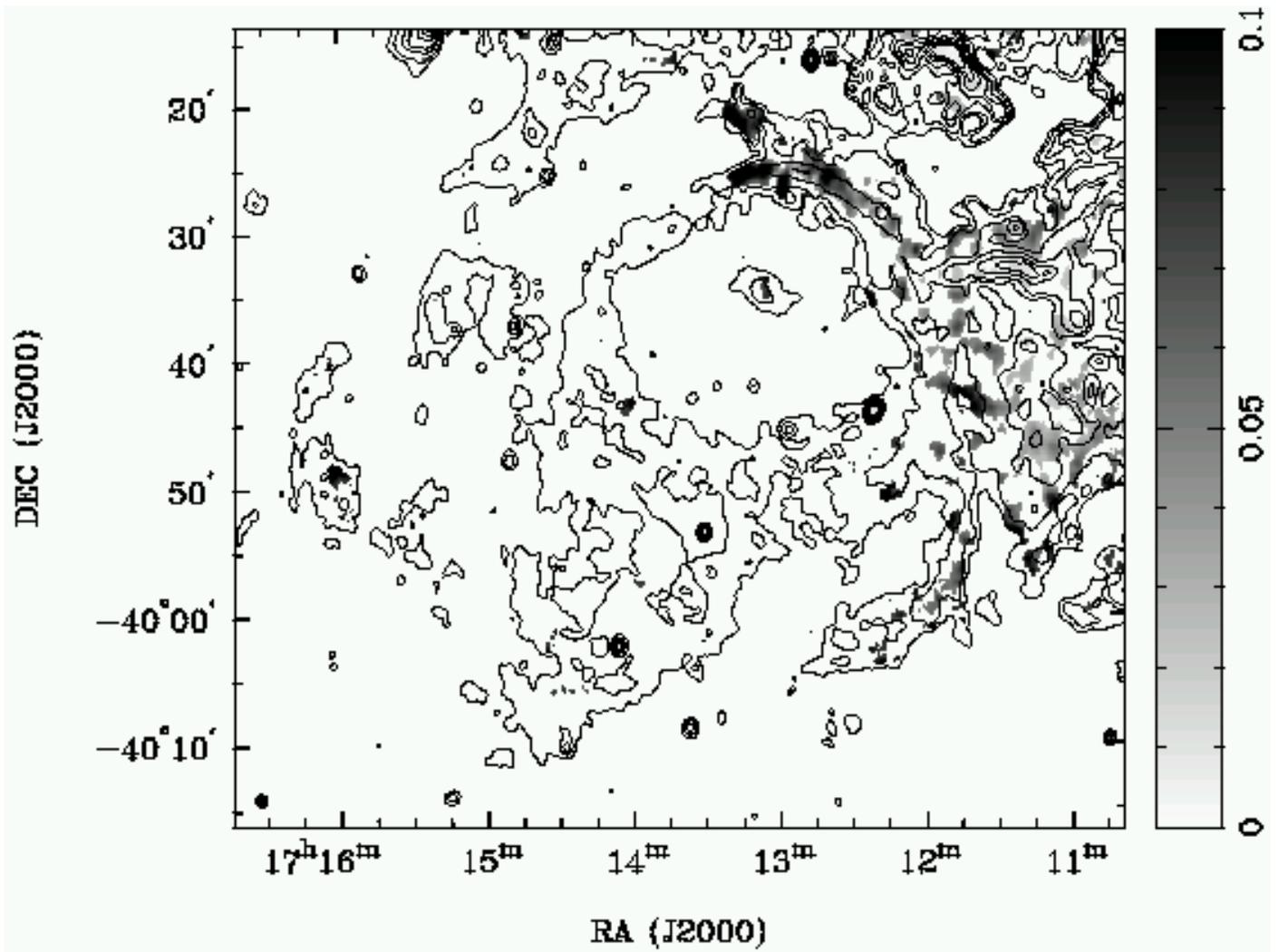}
\caption{Fractional polarization for G347.3--0.5 at 1.4~GHz overlaid
with the total intensity image over the whole SNR region.  
The grayscale is saturated to show regions
with low values and units are of fractional polarization.  The total
intensity contours are: 0.5, 1, 2, 3, 5, 7, 10, 30 $\times$ 8\mjb .
At the location of Arc~1,  the magnetic
field appears to be most ordered and the polarized intensity follows
the total intensity in this region.}
\label{fig-pol}
\end{figure}

We found significant linear polarization towards the western shell of
G347.3--0.5. In Figure~\ref{fig-pol} we show the fractional
polarization towards the SNR at 1.4~GHz. The mean fractional
polarization is $\sim 5$\% at 1.4~GHz and $\sim 10\%$ at 2.4~GHz,  
calculated by dividing the sum of
polarized intensity  by  the sum of total intensity.  The polarized
intensity is strongest  towards Arc~1, up to 12\% at 1.4~GHz and 30\%
at 2.5~GHz,  where it clearly
follows the arc-like distribution of the total intensity.  There is
weaker ($< 6$\%), diffuse polarized emission  from Arc~2. The
polarized emission towards the rest of the western SNR region 
is patchy and diffuse, with no
particular correlation with the radio continuum,  implying that the
magnetic field is not very highly ordered there and that most of  the
remnant has a high degree of depolarization. 
There is almost no polarization detected towards the
rest of the SNR, with exception of a few small regions, e.g.,  towards the
center region and the eastern SNR shell.

The multi-channel continuum capability of the ATCA can be used to
derive the rotation measure (RM) across the observing bandwidth,
caused by the Faraday rotation.  Using the frequencies within 1.4
and 2.4~GHz bandwidths, we get RM values towards Arc~1  of  
$> +100$ rad~m$^{-2}$. Our observed uncertainty in RM, $\Delta
{\rm RM}\sim 20$~rad~m$^{-2}$, implies an uncertainty in the intrinsic
polarization position angle $\Delta \phi_0 \sim \Delta {\rm RM} c^2
/\nu^2 \sim 1$~rad, when extrapolated from the RM measured within each
band. Furthermore, we are unable to calculate the RM between the 1.4 and
2.4~GHz bands, because these observations have different wavelengths and
angular resolution, and are thus subject to very different depolarization
and Faraday rotation effects by foreground ionized gas 
\citep[e.g.,][]{gaensler01}.  Thus we are unable to determine the
intrinsic orientation of polarization from the available data.

\section{Spectral Variations Across G347.3--0.5}

Recently, \citet{uchiyama03} presented a study of {\em
Chandra} observations towards the northwestern rim of G347.3--0.5.
In their spectral analysis of
individual spectra from field~1 they infer that the 
photon index is same in diffuse and bright SNR
regions. However, for their spectral fitting they used a blank sky
observations for the background subtraction and varying column
density for diffuse and bright regions. We also tried using blank sky
observations for the background subtraction and found no difference in
 spectral fit results from the fitting using 
 the background region from field~1. We also
found that leaving column density as a free parameter for each
individual spectrum results in very close photon index values. 
 As explained in
 an earlier section ($\S$~\ref{sec-xray}), the possible 
variation in column density across the northwestern shell of 
the SNR  seems unlikely; our use of a uniform value for $N_{\rm H}$
 accounts for the difference in spectral indices
 between our work and that of \citet{uchiyama03}.

We produced two summed spectra, one for the bright and one for the 
faint SNR regions in field~1,  which we fitted with 
 power law and {\tt SRCUT} models. The results from the fit are
summarized in Table~\ref{tab-joint-fit}.
The spectrum from bright regions has $\Gamma \sim 2.10$, while
 the spectrum from the faint regions has $\Gamma \sim 2.53$. 
When fitting the {\tt SRCUT} model we froze again the radio spectral index to
 the global remnant value of 0.6 and left the 
normalization as a free parameters.
 The resuting rolloff frequency for the bright
regions was $\nu_{rolloff} \sim 6.3\times 10^{17}$~Hz and for the faint
regions was $\nu_{rolloff} \sim 1.2\times 10^{17}$~Hz. 
The radio flux at 1~GHz estimated by {\tt SRCUT} of
$S_{\nu} \sim 4.2$~Jy for the bright regions and 
$S_{\nu} \sim 2.9$~Jy for the faint regions is broadly consistent 
with that measured from our 
radio data, which gives us confidence in this approach. 

 These results, as well as those from fitting the individual spectra,
 imply  that the fainter SNR
 regions have a steeper photon index and lower rolloff frequency 
than the bright regions. 
 Thus, our data suggest that acceleration of particles to
 higher energies is more efficient in the brighter regions, 
 perhaps because of the higher magnetic field in those
 regions. Another possibility is that electrons in the fainter 
 regions are older, i.e., they could be produced in bright regions 
 and then lose energy diffusing away from their origin.


\section{High-resolution Comparison of X-ray and Radio Data}

\begin{figure}
\plotone{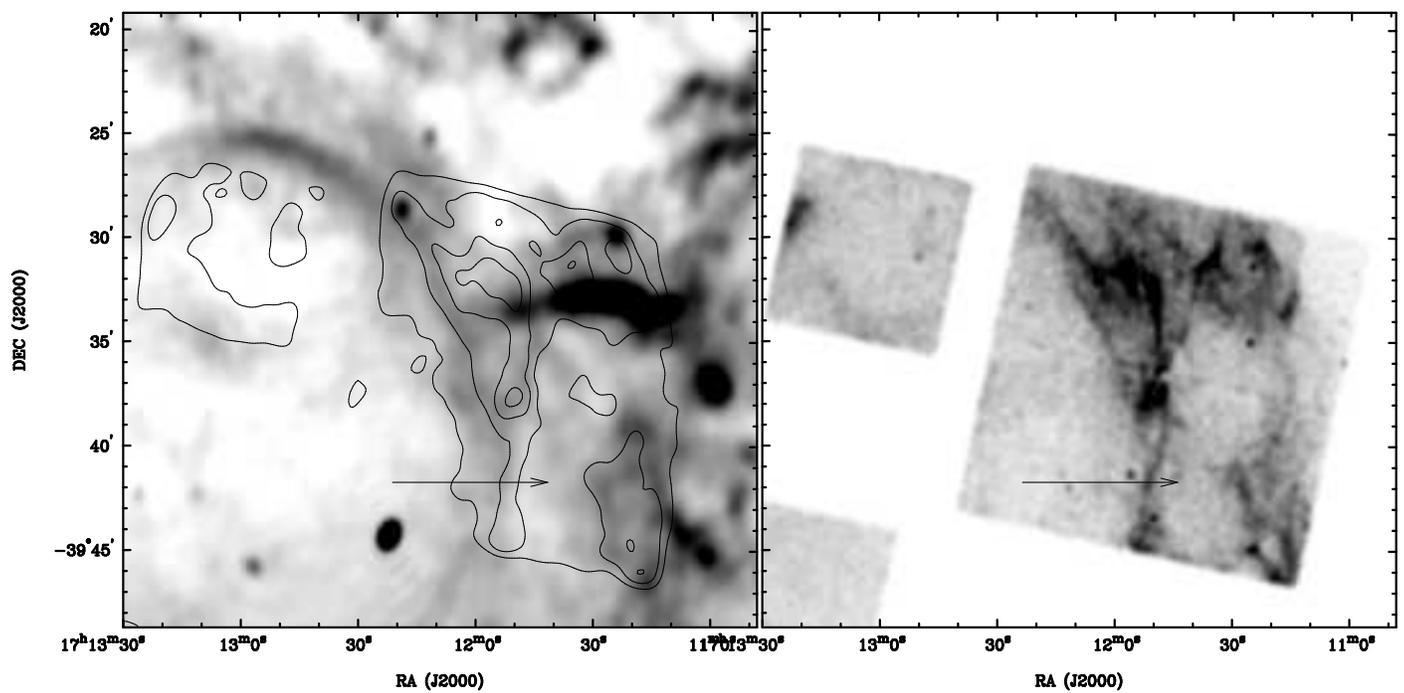}
\caption{ ({\em left}) 1.4~GHz greyscale image of the nortwestern SNR region
 overlaid with the {\em Chandra} contours convolved to the resolution
 of 1.4~GHz image of 47\arcsec. Contour levels are: 1, 2, 3 and 4 
\ee{-5}~counts\,arcsec$^2$\,s$^{-1}$. ({\em right}) 2\arcsec-resolution
 {\em Chandra} image of the  nortwestern SNR region in the right
 panel. In both panels the arrow  marks a profile cut region shown 
on Figure~\ref{fig-profile}.}
\label{fig-chandra+atca}
\end{figure}

\begin{figure}
\plotone{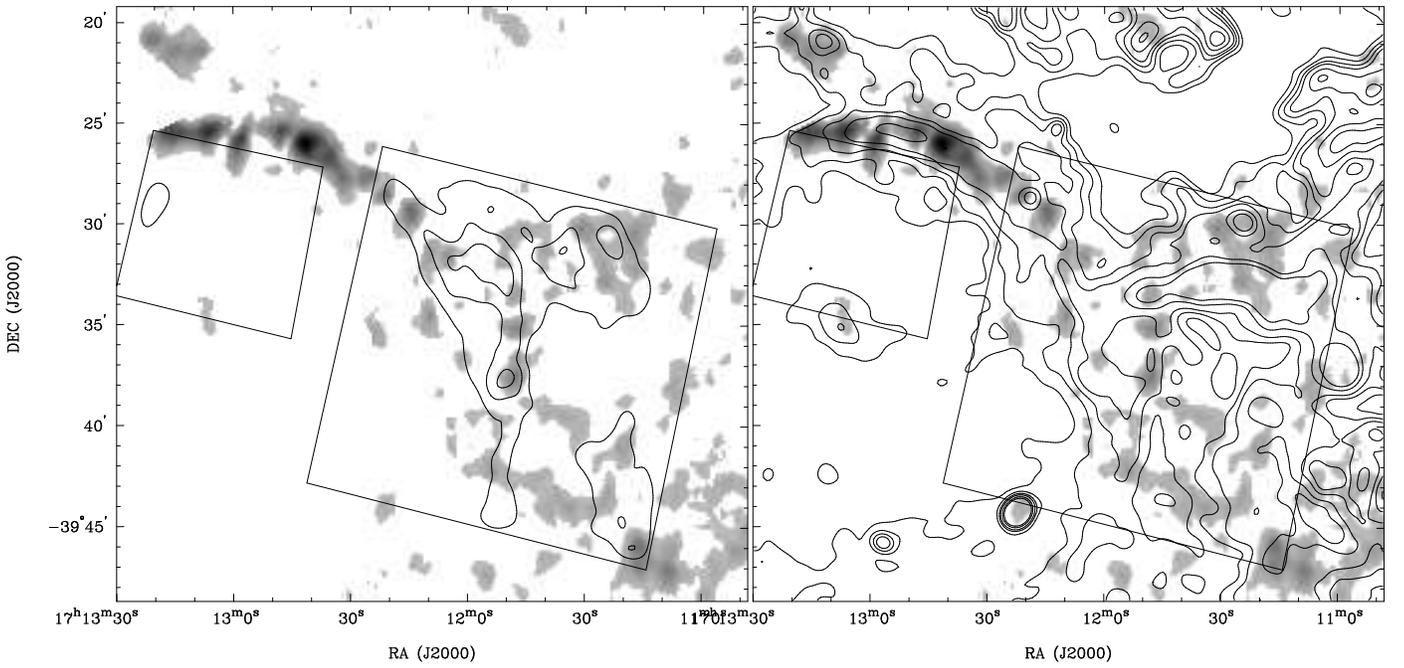}
\caption{1.4~GHz greyscale image of the polarized
 intensity towards the nortwestern SNR region
 overlaid in the left
 panel with the {\em Chandra} contours (convolved 
 to the resolution  of 1.4~GHz image of 47\arcsec) at 
 2, 3 and 4 \ee{-5}~counts\,arcsec$^2$\,s$^{-1}$ levels, 
and with the 1.4~GHz continuum contours at 0.5, 1, 1.5, 2, 3 and 4
 \ee{-2}~Jy. The squares are outlining the edges of the {\em Chandra} 
 detectors. }
\label{fig-pi}
\end{figure}

  Using our high-resolution radio  and X-ray images  we are able  for
 the first time to address the  correlation between the  radio and
 X-ray morphology in a great detail.  The {\em Chandra} and ATCA
 images of field~1 are overlaid in Figure~\ref{fig-chandra+atca}
 (left panel); we do not compare closely emission in field 2 because
  the radio emission is extremely weak there and shows no obvious
 structure.  The {\em Chandra} image of the field~1 is shown smoothed
 to resolution  of 2\arcsec\  (right panel), as well as convolved to
 the resolution  of 1.4~GHz ATCA image of 47\arcsec\ (contours in the
 left panel).  If the same population of electrons is responsible for
 the  synchrotron emission in both the radio and X-ray band,   one
 would expect similar morphology in these two bands, as appears to be
 the case for  SN~1006 \citep{allen01,dyer01}.  The correlation
 between the radio and X-ray images of G347.3--0.5 is not perfect,  as
 noted from the {\em ASCA} data \citep{slane99}.  Most notable is the
 lack of X-ray emission along the well-defined inner radio ring  and
 the lack of radio emission from the southwestern peak present in the
 X-ray image --- these regions were not imaged by us with {\em
 Chandra} so  we compare ATCA and {\em ROSAT} images in
 Figure~\ref{fig-atca}.  However, the broad  morphological agreement
 between the nonthermal X-ray and radio emission from field~1, and
 Arc~1 in particular  (Figure~\ref{fig-chandra+atca}), provides strong
 evidence that efficient particle acceleration is occurring in the
 northwestern part of the SNR.  Arc~2 shows mostly diffuse emission,
 but a bright  filament is present in the radio image which does not
 have a counterpart in the X-ray image.  The nonthermal radio nature
 of Arc~2 was questioned by  \citet{ellison01}  due to its proximity
 to a region of thermal emission from the adjacent \hr\ region.
 Indeed,  we find that this region has a large uncertainty in the
 radio spectral index, which implies confusion by the surrounding
 material. The X-ray spectrum of Arc~2 is clearly nonthermal; we found
 no indications of thermal X-ray emission towards this region.  
 Furthermore, the
 polarized intensity in this region, shown in Figure~\ref{fig-pi},
 shows  emission from Arc~2 which  morphologically resembles X-ray
 data rather than the radio data.   That implies that Arc~2's radio
 emission is at least partially nonthermal.

The expected relation of radio and synchrotron X-ray morphology in an
evolved, complex object like G347.3--0.5 is not simple.  While we
expect radio and X-ray emitting electrons to be born in the same shock
waves, their subsequent evolution can vary greatly, because of
different diffusive properties and the factor of 30,000 greater
radiative lifetime for a 1~GHz-radiating electron compared to a 
1~keV-radiating one.  In fact, an electron radiating its peak
synchrotron power in 1 keV photons has a half-life against synchrotron
losses of
\begin{equation}
t_{1/2} = 5.0 \times 10^4 B_{\mu{\rm G}}^{-3/2}
            \left( {h \nu \over {1 \ {\rm keV}}} \right)^{-1/2} 
            \ {\rm yr}
\end{equation}
where we have averaged over electron pitch angles.  (Losses to
inverse-Compton scattering from the cosmic microwave background (CMB)
dominate for $B < 3.2 \ \mu$G, the field strength with the energy
density of the CMB, so the maximum lifetime is about 8600 years for
1-keV-emitting electrons.) In a simple single shock wave, we might
then expect radio and X-ray synchrotron emission to appear at the same
location, with X-ray emission disappearing a shorter distance behind
the shock.  Such a morphology is evident in SN~1006 \citep{long03}
and in parts of RCW~86 \citep{rho02}.  If the magnetic field is
quite inhomogeneous, as we propose below, the lifetime of an
individual electron will also depend strongly on the mean magnetic
field along its particular diffusive trajectory.  Furthermore, weak
shocks may accelerate electrons to radio-emitting but not
X-ray-emitting energies, so we should not be surprised to find radio
emission without X-ray counterparts.  Finally, X-rays can be absorbed
by intervening gas at much lower column densities than would be
required for free-free absorption of radio emission.  In general,
while we would hope to find at least some shock-like (thin linear)
structures in X-rays and radio, there may not be an extremely close
correspondence overall between the two bands.

\begin{figure}
\plotone{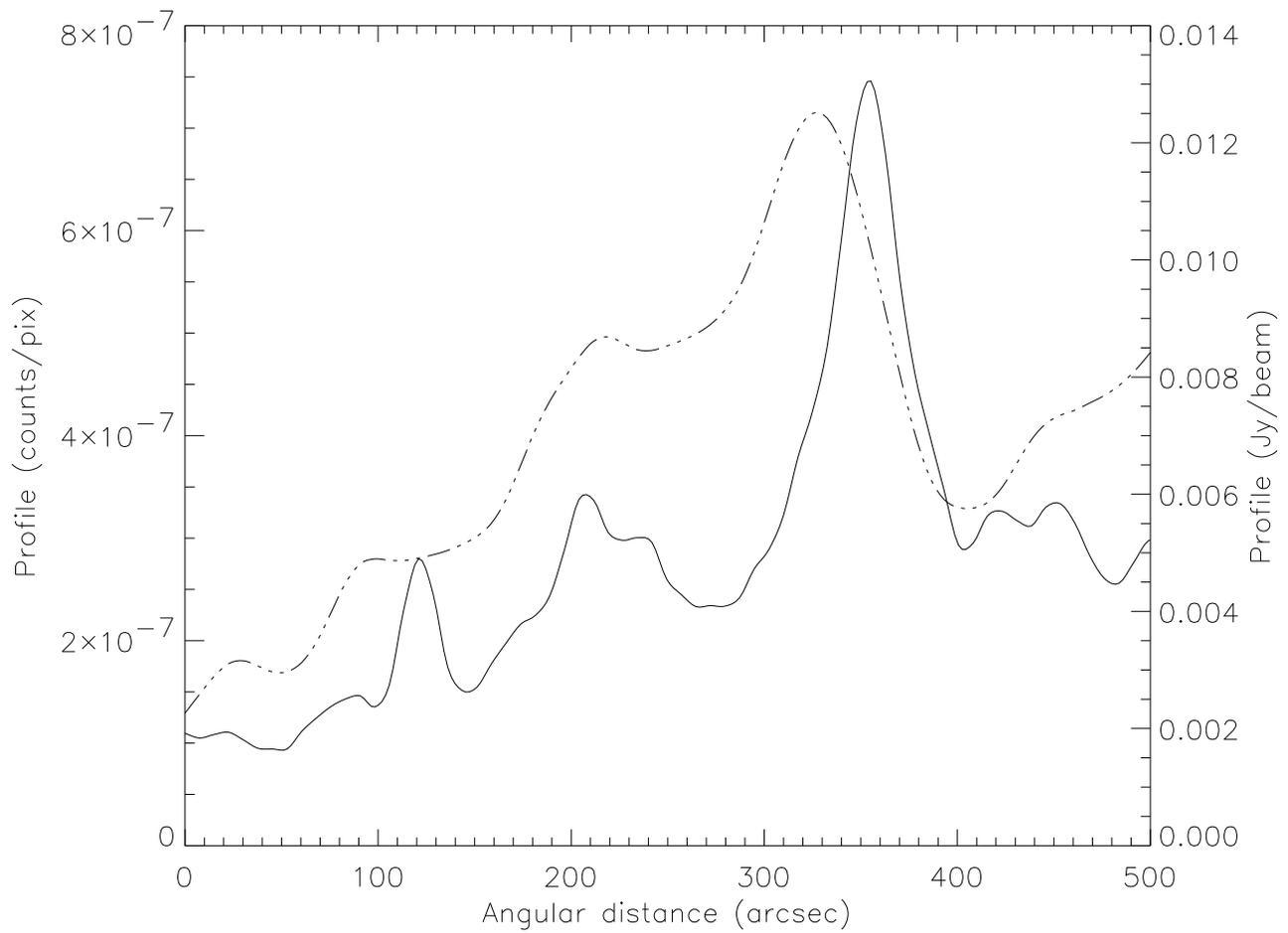}
\caption{Profile cut through Arc~1 from 1.4~GHz data (dash-dotted
line) and {\em Chandra} data (solid line) marked in 
Figure~\ref{fig-chandra+atca}.}
\label{fig-profile}
\end{figure}

Figure~\ref{fig-profile} shows profiles through the most prominent linear
feature seen in both radio and X-ray, a thin filament at the inner
edge  of Arc~1.  It is apparent that the turn-on of radio and X-rays is
nearly coincident (we recall that the resolution of the radio image is
47\arcsec), while the radio emission extends considerably further than the
X-rays.  The width of the X-ray filament is about 40\arcsec\ (1.2~pc
at a distance of 6~kpc).  If the leading edge of the structure is in
fact a shock seen close to edge on, the X-ray width could result from
synchrotron losses in a $3 \ \mu$G field if the relativistic electrons
are convected downstream (i.e., in the plane of the sky) at the modest
velocity of about 140~km\,s$^{-1}$.  As we argue below, the broadband
spectrum of G347.3--0.5 can be explained if the magnetic-field
structure consists of small regions with $B_1 \sim 15 \ \mu$G
occupying about 1\% of the volume, with a much lower field 
($B_2 < 1 \mu$G) occupying the rest.  The mean field sampled by an
electron would be fairly close to $B_2$, in this case, unless
electrons were somehow trapped in the high-field regions.  We
therefore expect lifetimes of 1-keV-emitting electrons to be near the
CMB limit of about 9000~yr, though in particular regions they could be
much shorter.  (An electron living all its life in $B_1$ would last
only about 140~yr).

 
\section{Origin of TeV Emission from G347.3--0.5}

 An earlier broad-band emission model for G347.3--0.5 by \citet{ellison01}
 implied that the IC mechanism was responsible for producing TeV
 emission from the remnant.  The derived maximum electron energy of
 $E_{max} \sim 14 \times 10^{12}$~eV falls below the ``knee'' energy,  but
 they showed that in addition to efficient acceleration of electrons
 there can be efficient  acceleration of ions (Fe$^{+26}$ in
 particular), whose maximum energy could reach the desired energy of
 $\sim 10^{15}$~eV.  However, for this study only a single TeV flux
 measurement from the 3.8-m CANGAROO telescope was available. Most
 recently, \citet{enomoto02} published a $\gamma$-ray spectrum for
 G347.3--0.5  obtained with the new 10-m telescope, and suggested that
 the IC mechanism cannot produce the  observed shape of the TeV
 spectrum. They modeled the TeV spectrum assuming pion-decay,  which
 implies a higher particle density for the ambient medium than allowed
 by the X-ray emitting gas inside the SNR \citep{slane99}. If TeV emission is
 produced in regions outside the boundary of the SNR shock,  where
 particles are upscattered through  the diffusive acceleration
 process,  the density values derived by \citet{enomoto02} from the
 pion-decay model  might be plausible since the molecular cloud is so
 close to the remnant. In this case  the bulk of TeV emission would
 have to originate in the cloud, somewhere beyond the SNR shell. Their
 pion-decay spectrum is inconsistent with EGRET observations
 \citep{butt01,reimer02}, unless the proton spectrum has  a cutoff
 below some energy, as also noted by  \citet{uchiyama03}.  This
 would result naturally if the protons must diffuse ahead of the shock
 to the molecular cloud, and if the diffusion coefficient increases
 with proton energy.   In such a scenario only the highest-energy
 protons would reach the target,  so that the pion-decay model would
 require an even higher target (molecular cloud) density, and demand
 even greater production efficiency of cosmic-ray protons  at the
 shock. \citet{uchiyama03} applied a two-zone model for the 
acceleration and diffusion of particles in an effort to 
explain the irregular morphology along the
SNR limb.  They considered the bright SNR
 filaments as locations where particles are accelerated, and the 
fainter regions where the electrons diffuse to after being accelerated. 
For a distance of 6 kpc, they derive a field strength of
50\,$\mu$G,  but require that this field be similar in the
filamentary and diffuse emission regions. This then implies that
 the spectral cutoff in the diffuse emission regions
is due to radiative losses while that in the filmants is due to
diffusive escape of the particles. The resulting IC emission falls
below the observed CANGAROO flux, thus requiering some other mechanism
to produce the TeV emission. The IC mechanism was also rejected in the
 most recent study of {\em ASCA} data by \citet{pannuti03}, 
because it required  unrealistically low magnetic field 
filling factor (see next section). We wished to see, however, if our new
{\it Chandra} data really supported ruling out the IC mechanism
for the TeV emission.

\subsection{The Broad-band Model}

\begin{figure}
\epsscale{1.0}
\plotone{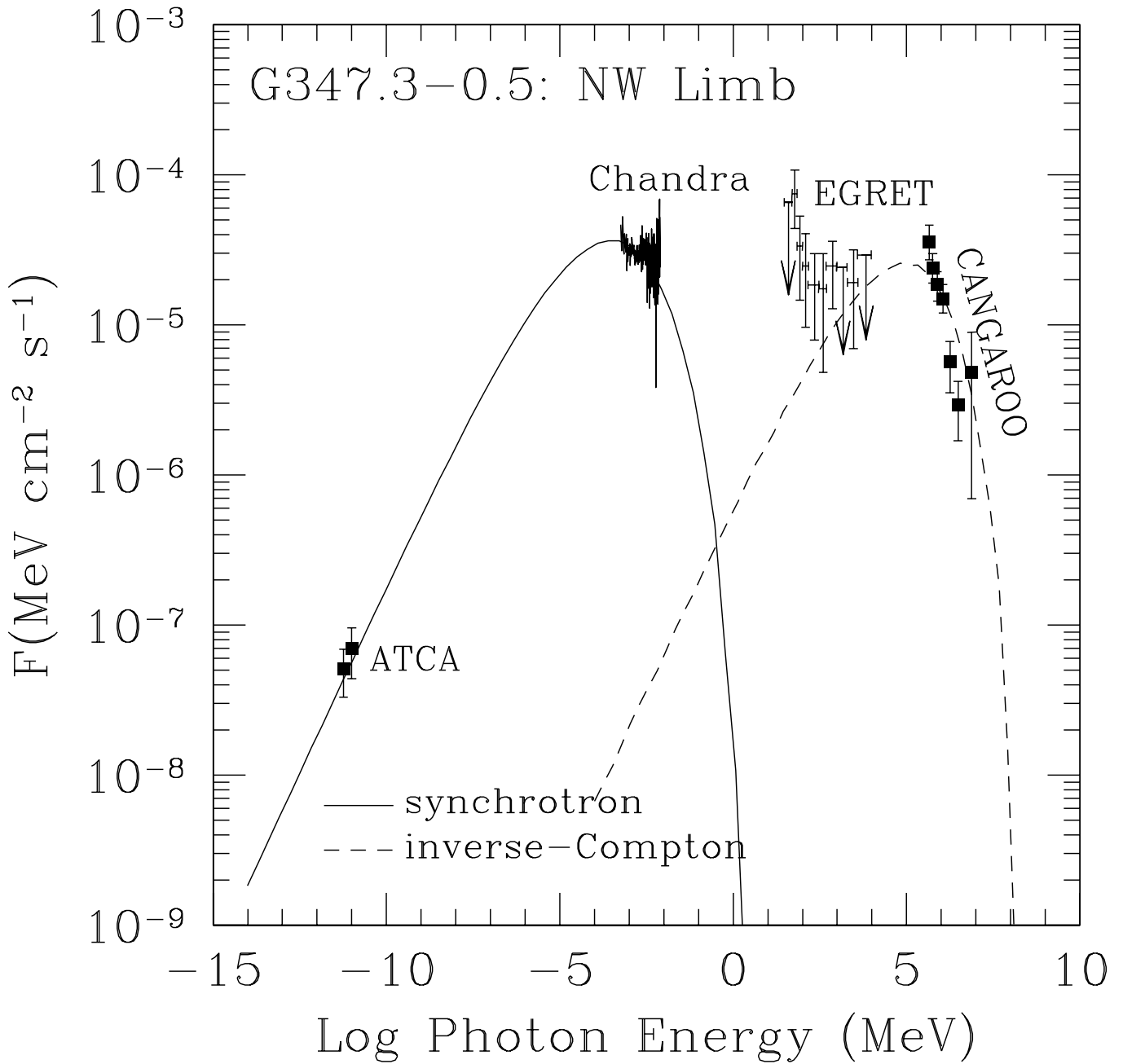}
\caption{Model of the broad-band emission from G347.3--0.5.   The
 model shows estimates of the photon spectra associated with
 synchrotron and IC emission (see text).}
\label{fig-model}
\end{figure}

In Figure~\ref{fig-model} we show measurements used to model the
broad-band emission from G347.3--0.5.  While the CANGAROO
 observations have poor angular resolution \citep[for G347.3--0.5
observations estimated to be 0\fdg 23;][]{enomoto02}, we 
 shall assume that the
 TeV emission originates from particles in a broad region in the
 northwest of the SNR \citep[see][]{muraishi00},  
 which  includes the synchrotron-emitting
 regions. The similar radio and X-ray morphology towards the
 northwest suggests  a common origin for the synchrotron-emitting
 particles at least in this part of G347.3-0.5. 
To match the spatial resolution of the CANGAROO data  
we produced a summed {\em Chandra} spectrum for the region covered
by ACIS-I  CCDs in field~1. The resulting unabsorbed X-ray flux 
 in the 0.5--10.0~keV band is $\sim 1.3$\ee{-10}\erg.
To measure radio flux densities, we convolved the radio images
to 60\arcsec, which are then corrected for variations in the
background level determined from the average of several areas around
the region of interest.  We obtained an integrated flux density from
the region corresponding to field 1
 of $6.7 \pm 2.0$ Jy at 1.4~GHz, and  $5.6 \pm 1.7$ Jy at 2.4~GHz. 
The  EGRET measurements are taken from
\citet{reimer02}, which used flux values from \citet{hartman99}.   The
EGRET values correspond to the nearby source 3EG~J1714--3857, which is
either associated with G347.3--0.5 or else provides an upper limit to
any such emission from the SNR. The CANGAROO measurements are taken
from \citet{enomoto02}.

We model the synchrotron and IC  emission from G347.3--0.5 using a
power law energy distribution modified by an
exponential cutoff \citep[e.g.,][]{gaisser98}: 
\begin{equation}
\frac{dN}{dE} = A_e~E^{- \sigma} exp \left [- 
 \left ( \frac{E}{E_{max}} \right ) ^{\alpha_{path}} \right ] ,
\end{equation} 
where $A_e$ is the normalization factor, $\sigma$ is the 
index of the electron distribution and $E_{max}$ is
the maximum energy of accelerated particles.
 We included additional parameter, $\alpha_{path}$, in the
exponential function. This is a phenomenological parameter that allows for 
a broadening of cutoff, which accounts for 
 variations in magnetic field from place to place, 
or in $E_{max}$, or both. A similar model was also
used by  \citet{uchiyama03}, or in a different form by \citet{ellison01}. 
We use a steady-state model which assumes that the synchrotron 
and IC emission comes from a single population of relativistic 
electrons.  Details of the model are given in Appendix~A.  
The model yields estimates of the photon  spectra produced 
by synchrotron and IC emission mechanisms in
the $\nu F_{\nu}$ format:
\begin{equation}
\nu F_{\nu} (synch) = \frac{V_{E}}{4\pi D^{2}_{pc}}~f_B~
 \nu~P_{tot}(\nu, E), 
\label{eq-synch}
\end{equation}
\begin{equation}
\nu F_{\nu} (IC) = \frac{V_{E}}{4\pi D^{2}_{pc}}~
 E^2~\frac{dn_{\gamma}(\varepsilon_{\gamma})}{dt},
\end{equation}
where $D_{pc}$ is the distance to the SNR in pc,  $V_{E}$ is the
electron emission volume (assuming a thin spherical shell geometry)
given as \citep{ellison01}: 
\begin{equation}
V_{E} = f_{E}~\frac{4 \pi}{3} (R_{O}^{3} - R_{I}^3 ).
\end{equation}
$R_{O}$ and $R_{I}$ are the outer and inner shock radius respectively,
and $f_{E}$ represents the fraction of  the shell volume producing the
emission seen from the northwestern SNR limb. The magnetic field
filling factor $f_B$ in equation (\ref{eq-synch}) 
corresponds to the fraction of the volume of
IC-emitting electrons containing the magnetic field responsible for the
synchrotron emission \citep[e.g.,][]{allen01}.  $P_{tot}(\nu, E)$ and
$\frac{dn_{\gamma}(\varepsilon_{\gamma})}{dt}$ are synchrotron and IC
emissivities and are given in Appendix A.

Since the magnetic field strength $B$ and maximum particle energy
$E_{max}$ cannot be determined independently from the synchrotron
spectrum, we first modeled the CANGAROO data
with an IC spectrum (treating the EGRET data as an upper  limit for
the flux detectable from the SNR) which depends only on $E_{max}$ (see
 Appendix A). In this way we determined the
parameters for the electron number distribution:  maximum electron
energy $E_{max}= 5$~TeV and  particle index $\sigma = 2.0$.
We then varied the magnetic field strength, which is fixed by the
 ratio of peak  emission frequencies $R_{\nu}$.
An estimate good to about 25\% is given by (see Appendix B)
\begin{equation}
B=9 \times 10^4 C_{\rm path}^{-1} R_{\nu}^{-1} {\rm G}.
\end{equation}
The parameter $C_{\rm path}$
depends on the radio spectral index and on the parameter 
$\alpha_{\rm path}$ describing the breadth of the cutoff in the
electron spectrum; by experimentation we found that for $\sigma = 2,$
a value of $\alpha_{\rm path} = 0.5$ described the X-ray data well.
For these values, $C_{\rm path} = 16$ which, with an estimate
for $R_\nu \sim 10^{8.5}$, gives $B \sim 20 \ \mu$G.
We used our full model calculation for a more precise assessment
and found that in order to match the shape of the X-ray spectrum we
require $B \sim 15\,\mu$G. 
However, this would overpredict the
synchrotron  portion of the spectrum by a factor of 100. It is this
problem that caused \citet{enomoto02} to rule out a synchrotron/IC
model.  But we can rescue such models by assuming that the magnetic
field fills only a portion of the volume  occupied by the IC-emitting
electrons, and take a value of $\sim 0.01$ for the magnetic
field filling factor $f_B$. This filling factor is given by (see
Appendix B):
\begin{equation}
f_B=C(\sigma) R_{S} B^{(\sigma+1)/2}
\end{equation}
 where $R_S$ is the ratio between synchrotron and IC flux at a frequency in the
power-law part of both spectra, and the values of $C(\sigma)$ are given in
Appendix~B. The values of the parameters
used in the model are summarized in Table~\ref{tab-model}. 

We note that using the {\tt SRCUT}-type model 
(i.e., setting $\alpha_{path}=
1$), our model gives parameters for the broad-band spectrum
consistent with those of \citet{pannuti03}, who derived high
magnetic field ($\sim 150~\mu$G) and unrealistically low
magnetic field filling factor ($\sim 10^{-3}$),  the latter being
the reason they rejected IC as a plausible
mechanism for TeV emission production in G347.3--0.5. However, we find
that the X-ray data seem to prefer slightly slower steepening of the
cutoff than the sharp exponential cutoff.  Thus our model with
$\alpha_{path}= 0.5$ results in  different broad-band model parameters
which are more  reasonable from the physical point of view.
Furthermore, our results illustrate the strong dependence  of inferred
parameters on the detailed model used to describe the  synchrotron
X-rays. This strong dependence means that particular parameter values
obtained from this kind of broad-band fitting may be far less
well-determined than their statistical uncertainties within a
particular model may seem to indicate.  Since our purpose here is
mainly to illustrate that a synchrotron/IC broadband model is not ruled out by
the data, we need only demonstrate that some reasonable model in this
class can reproduce the observations, and we feel we have done so.

Thus, we find that, if the CANGAROO spectrum of G347.3--0.5 is
correct, we have to decouple the electron emitting volume from the
magnetic field volume to model the synchrotron spectrum with the
parameters from the IC spectrum.  
We note that a magnetic field filling factor ($f_B$) value 
of $\sim 1$\% seems very low, but MHD simulations do predict magnetic
 field enhancements in small-scale features produced 
by turbulence generated in SNR shocks \citep{jun99}. This is supported
by the radio polarization maps of G348.3-0.5, which show 
that the magnetic filed in this SNR is very
patchy, which could be, in part,  due to the turbulent amplification.
  Also, this is the factor by which the volume occupied by
magnetic field producing the bulk of the synchrotron emission is
smaller than the volume occupied by ultrarelativistic electrons
producing IC photons. In other words, the $15\ \mu$G magnetic field is
confined to $\sim 1$\% of the region from which the IC emission is
emitted. This requires that the rest of the region has much lower
magnetic field, $\sim 0.7\,\mu$G, in order not to exceed 
the total observed synchrotron emission from the SNR.

Our value for magnetic field is consistent with that calculated by  
\citet{ellison01}, but our value for maximum energy of accelerated
particles is factor of 3 lower. This is not surprising since
$E_{max}$ is constrained by the TeV spectrum and their model was based
on a single TeV measurement, which lies in the middle of the TeV
spectrum and would thus push $E_{max}$ to the higher values. Using
the rolloff frequency obtained from the {\tt SRCUT} fit to field~1
spectra and our value for magnetic field, we get $E_{max}\sim
5.0$~TeV, using relationship 
\begin{equation}
\nu_{rolloff}\approx 1.61 \times 10^{16} \left ( \frac{B}{10
\mu {\rm G}} \right ) \left ( \frac{E}{10 {\rm TeV}} \right )^2 ~{\rm
Hz,}
\end{equation}
which is the corrected version of the relationship reported in 
\citet{reynolds99} which contained a confusion of a factor of 0.29 between
maximum and characteristic frequencies \citep[see e.g.,][] {pacholczyk70}. 
This value of $E_{max}$ is thus consistent with our model. In any case,
given the uncertainties inherent in this kind of broad-band modeling
of a highly inhomogeneous source, even a factor of 3 difference in
inferred parameters should be regarded as rough consistency.


\subsection{Maximum Energy of Accelerated Electrons}

Our fitted value of $E_{max}$ of 5~TeV, with a magnetic field strength of
$15 \ \mu$G, gives a peak emitting frequency of $\nu_{ms} \sim 2
\times 10^{15}$ Hz.  This frequency is low enough to put significant
constraints on shock-acceleration models producing the electron
spectral cutoff by radiative losses. 
Simple estimates of the cutoff energy \citep[e.g.,][]{reynolds98} 
in which the electron scattering mean free path is a
constant factor $\eta \ge 1$ times the gyroradius give
\begin{equation}
E_{max} \sim 20 \ \eta^{-1/2}
      \left( B \over {15 \ \mu{\rm G}} \right)^{-1/2} u_8 \ {\rm TeV}
\end{equation}
where the shock normal is parallel to the mean upstream magnetic field,
and 
\begin{equation}
E_{max} \sim 20 \ \eta^{+1/2}
      \left( B \over {15 \ \mu{\rm G}} \right)^{-1/2} u_8 \ {\rm TeV}
\end{equation}
where the shock normal is more nearly perpendicular.  The only difference
is in the $\eta$-dependence, 
due to more rapid acceleration in perpendicular shocks
\citep{jokipii87}. Also, $u_8$ is the shock velocity in units of $10^8$
cm s$^{-1}$, and $B$ is the postshock field strength.
The above estimate is weakly dependent on the
assumption that the shock compression ratio $r$ is 4.  

These expressions give a cutoff frequency
\begin{equation}
\nu_{ms}({\rm perp}) \equiv 1.82 \times 10^{18} E_{max}^2 B 
    = 3 \times 10^{16} \eta~u_8^2 \ {\rm Hz}
\end{equation}
for a perpendicular shock, and
\begin{equation}
\nu_{ms}({\rm par}) = 3 \times 10^{16} \eta^{-1}~u_8^2 \ {\rm Hz}
\end{equation}
for a parallel shock.
These frequencies are independent of the magnetic field strength,
since $E_{max} \propto B^{-1/2}$

If the shock velocity in G347.3--0.5 is about 1000 km s$^{-1}$
\citep{ellison01}, it is impossible to produce the observed
low cutoff frequency in perpendicular shocks --- the acceleration
rate is too fast.  If the shocks are parallel, larger $\eta$ will
slow the acceleration rate and lower the cutoff frequency.  For
instance, if $u_{\rm sh} = 900$ km s$^{-1}$, then $\eta \sim 10$
would produce our inferred cutoff frequency.  In the quasilinear
approximation, electron scattering is due to the amplitude of
magnetic fluctuations $\delta B$ with wavelengths comparable to
the electron's gyroradius.  In this picture, $\eta = (\delta B/B)^{-2}$.
In strong turbulence, $\eta$ approaches 1 (Bohm limit).
A value of 10 is often assumed for moderately strong ($\delta B/B
\sim 0.3$) turbulence.

The issue of the nature of the cutoff has important implications for
the acceleration of ions.  If the electron spectrum is limited by
radiative losses, the proton spectrum might extend to much higher
energies, while any other cutoff mechanism for electrons should affect
protons as well.  We deduce that if the shocks accelerating the
electrons producing synchrotron X-rays in G347.3--0.5 are largely
parallel shocks with fairly high levels of MHD turbulence, then the
rate of electron acceleration can balance the loss rate for electron
energies around the 5~TeV cutoff we obtain from fitting the TeV photon
spectrum, and we cannot constrain any turnover in the accelerated
proton spectrum.  If the shocks do not have these properties, or if the
acceleration is due to some other process altogether, we should expect
that the proton spectrum also cuts off around 5~TeV. 


Thus, assuming that protons are accelerated in the same manner as
electrons, we can use the same particle distribution for electrons and
protons (i.e., $\sigma=\sigma_p$ and $E_{max}=E_{max,p}$) to
calculate total energy content of the relativistic electrons and
protons \citep[see][]{allen01}.  We obtained a total electron energy
 of 3.7\ee{46}~erg and a total proton energy of
1.7\ee{49}~erg, which yields total particle energy that is
significantly smaller than the SNR energy budget ($\sim 1-2 \times
10^{51}$\,erg). The total magnetic energy  yields 4.6\ee{46}~erg
for  $B=15\mu$G and $f_B=0.01$, remarkably close to equipartition
with electrons.

\section{CONCLUSIONS}

{\em Chandra} and ATCA observations of G347.3--0.5 have been used
to investigate the SNR's morphology and spectral properties with 
high angular resolution. The main results are  summarized below.

The X-ray emission  from the remnant is dominated
by  nonthermal power law emission, as found in previous observations.
High resolution X-ray observations with {\em Chandra} reveal a complex
morphology of the northwestern SNR region composed of bright filaments
embedded in faint diffuse emission. Our spectral analysis implies that
there are significant variations between the spectral properties of
the fainter regions, which have steeper spectra ($\Gamma \sim 2.5$), 
and that of the bright regions, which have flatter spectra ($\Gamma
\sim 2.1$). 

To improve an estimate of the thermal emission in the SNR, we studied   
in particular the emission in the ACIS-S3 detectors which have the 
best response to the soft emission. We found a possible 
trace of thermal emission, but were unable to constrain its properties. 
 No thermal component was detected in regions covered by the 
ACIS-I detectors.


Due to a complex environment the SNR is located in, we were unable to
derive an accurate spectral index for radio emission.
Using the ATCA data at 1.4 and 2.5~GHz we derived an approximate 
spectral index of the northwestern SNR region to be $0.50\pm0.40$.  

Significant linear polarization of $\sim 5$--10\% is  detected
with the magnetic field being most ordered towards the northwestern
 SNR filament, where the linear polarization reaches 12--30\%.
The low mean value is comparable to that
found in the historical shell remnants 
\citep[see references in][]{reynolds93}, and is considerably smaller
than often found in larger, older remnants. It is thus consistent
with the otherwise surprising result that a remnant as large as
G347.3-0.5 could have such strong shock acceleration.


We used the small scale morphology of {\em Chandra} data to 
identify possible regions of particle acceleration in the SNR.
The X-ray morphology of the northwestern SNR region
corresponds fairly well with the radio morphology, implying that the
same population of electrons is responsible for the emission in both
bands.  
We show that synchrotron/IC models cannot be ruled out for the
broad-band spectrum, if we allow the possibility that the magnetic
field is spatially inhomogeneous and enhanced in small
regions. The data require a magnetic field of  $\sim 15 \mu$G occupying
$\sim 1\%$ of the IC-emitting volume which is filled with relativistic
electrons.  The maximum energy of accelerated electrons is found to be
$\sim 5$~TeV. The derived total electron and magnetic energies are close to
equipartition in this model. 

While non-thermal bremsstrahlung is not considered as a potential 
TeV emission mechanism because it would violate the  limits on 
the thermal part of the bremsstrahlung emission from the SNR 
\citep{ellison01,pannuti03}, there are 
objections for both IC and pion-decay processes. IC requires small
 magnetic field filling factor of 
nonthermal electrons ($f_B \sim 1\%$), while pion-decay requires extreme 
gas densities. While the very 
small filling factor may seem implausible, there 
is some observational evidence for such regions in G347.3--0.5.   
Future $\gamma$-ray telescopes with improved spatial resolution will
help localize the TeV emission regions more precisely, enabling more
accurate broad-band models to be derived.  

\acknowledgements

We thank G. Allen and D. Ellison for invaluable discussions   and
D. Galloway for help with programing in IDL.  This work was supported
in part by NASA contract NAS8--39073 (POS),  grant GO0--1123X (JSL)
and GO2--3080B (JPH).

The Australia Telescope Compact Array is part of the Australia
Telescope  funded by the Commonwealth of Australia for operation as  a
National Facility, managed by CSIRO.   This research has made use of
the NASA/IPAC Infrared Science Archive, which is operated by the Jet
Propulsion Laboratory, California Institute of Technology, under
contract with the National Aeronautics and Space Administration.

\appendix

\section{Synchrotron and Inverse Compton Models}

The energy distribution in our synchrotron and inverse
Compton (IC) models is taken to be a power law with an exponential
cutoff \citep[e.g][]{gaisser98}: 
\begin{equation}
\frac{dN}{dE} = A_e~E^{-\sigma} exp \left [- 
 \left ( \frac{E}{E_{max}} \right )^{\alpha_{path}} \right ] ,
\end{equation} 
where $A_e$ is the normalization factor, $\sigma$ is the index of the
 particle distribution $E_{max}$ is the maximum energy of
accelerated particles, and   $\alpha_{path}$  is an empirical 
parameter allowing a
broadening of the cutoff apparently required by observations
\citep[see e.g.,][]{ellison01,uchiyama03}. 

The total synchrotron power radiated by a single electron is
\citep[e.g.,][]{blum70} 
\begin{equation}
p(\nu) = \frac{\sqrt{3} e^3 B_{\bot}}{m c^2} \frac{\nu}{\nu_c}
\int^{\infty}_{\nu /\nu_c} K_{5/3}(\xi)d\xi ,
\end{equation}
where $\nu_c$ is the critical frequency given by  
\begin{equation}
\nu_c = \frac{3 e B_{\bot}}{4 \pi m^3 c^5} E^2,
\end{equation}
$B_{\bot}$ is the magnetic field component perpendicular to the line
of sight, and $K_{5/3}$  is the modified Bessel function of  5/3
order.  The total power radiated by the population of electrons with
the modified power law distribution $N(E) \equiv dN/dE$ is then 
\begin{equation}
P_{tot}(\nu ,E)=\int_{E_{min}}^{E_{max}} p(\nu) N(E) dE.
\end{equation}

For the IC spectrum we follow the  formulae of \citet{baring99}. The
 probability that a photon with initial energy $\varepsilon_s m_e c^2$
 will collide  with an electron of energy $\gamma_e=(E_{kin} - m_e
 c^2)/m_e c^2$  and upscatter to the  energy $\varepsilon_{\gamma} m_e
 c^2$ is given in the general case  with the Klein-Nishina cross
 section as: 
\begin{equation}
\sigma_{\rm K-N}(\varepsilon_s, \gamma_e; \varepsilon_{\gamma})=
\frac{2 \pi r_{0}^2}{\varepsilon_s \gamma_{e}^{2}} 
\left [ 2q~log_e q + 1 + q - 2q^2 + 
\frac{\delta_e^2 q^2 (1 - q)}{2(1 + \delta_e q)} \right ],
\end{equation}
where $r_0 = e^2/(m_e c^2)$ is the electron radius,  $\delta_e = 4
\varepsilon_s \gamma_e$ is a parameter that controls the importance of
photon recoil and Klein-Nishina effects \citep{blum70}, and $q$ is
given as: 
\begin{equation}
q = \frac{\varepsilon_{\gamma}}{\delta_e (\gamma_e - \varepsilon_{\gamma})},
~~~0\ge q \ge 1.
\end{equation}
We take the photon field to be the cosmic microwave background (CMB)
radiation, which has the spectral  distribution of a blackbody at
temperature of $T=2.73$~K: 
\begin{equation}
n_{\gamma} (\varepsilon_s)=\left ( \frac{\varepsilon_{s}^{2}}{\pi^2 \lambda_{c}^{3}} 
\right ) \left ( \frac{1}{e^{\varepsilon_s/\Theta} - 1} 
\right ) 
\end{equation} 
where $\Theta=\frac{k T}{m_e c^2}$, and $\lambda_c=\hbar /(m_e c)$ is
the Compton wavelength. The IC emissivity is then 
\begin{equation}
\frac{dn_{\gamma}(\varepsilon_{\gamma})}{dt}=
c \int N_e (\gamma_e) d\gamma_e \int n_{\gamma}(\varepsilon_s) d\varepsilon_s
\sigma_{\rm K-N}(\varepsilon_s, \gamma_e; \varepsilon_{\gamma}), 
\end{equation}
where $N_e (\gamma_e) \equiv dN/dE$ is the electron energy
distribution.  We have ignored other sources of seed photons for two
reasons.  First, in typical SNR environments the energy density of the
local IR/optical radiation field is an order of magnitude or less 
than the CMB \citep{gaisser98}.  While no emission in {\em
IRAS (Infrared Astronomical Satellite)}  images can be clearly
associated with G347.3--0.5, an upper limit for the flux at 100$\mu$m
can be estimated by taking all emission within the remnant boundaries.
This gives about 450 Jy, or a mean intensity of about 0.16 Jy
arcmin$^{-2}$, implying in the source an energy density $\nu u_\nu
\sim 0.01$ eV cm$^{-3}$.  This upper limit is far smaller than that of the CMB
(0.26 eV cm$^{-3}$).  Upper limits from the shorter-wavelength {\em
IRAS} bands are even smaller than this. Second, the Klein-Nishina parameter
$\delta_e$ is  $\sim 10^{-2}$ for our 5~TeV electrons, meaning that the
scattering cross-section is essentially the Thompson cross-section.
For $\delta_e > 1,$ the cross section begins to drop substantially due
to Klein-Nishina suppression.  This corresponds to photon wavelengths
shortward of about 50 $\mu$m.  So mid-to near-IR and optical seed
photons would contribute substantially less to the IC emission, even
if their  energy densities were comparable to that of the CMB.  We are
thus justified in neglecting IR and higher photon fields in
calculating the inverse-Compton emission from G347.3--0.5.


\section{Magnetic Field and Its Filling Factor}

For a simple homogeneous source filled with relativistic electrons and
magnetic field (with a filling factor $f_B$), we can write down simple
relations between the synchrotron emission and inverse Compton
emission from CMB seed photons.  We consider spectra plotted as $\nu
F_\nu$, so that both components rise with the same slope $1-\alpha$
(where $\alpha$ is radio spectral index) to peaks at frequencies
$\nu_{ps}$ and $\nu_{pi}$, respectively.  Since Compton upscattering
increases scattered photon energies by a factor $\sim 2 (E/m_e
c^2)^2$, a photon at the peak of the 2.73 K CMB spectrum with
frequency $5.6 \times 10^{10}$ Hz emerges at a frequency $\nu_i = 1.7
\times 10^{23} E^2$ Hz (with electron energy $E$ in erg).  Now
electrons with energy $E_m$ radiate the peak of their synchrotron
radiation (SR) spectrum at a frequency $\nu_{m} = 1.82 \times 10^{18}
E_m^2 B$ \citep[e.g.,][]{pacholczyk70}.  However, the peak of
the resulting $\nu F_\nu$ spectrum is not necessarily at $\nu_m$.
The delta-function approximation to the single-electron synchrotron
emissivity, integrated over our modified exponentially cut off
electron spectrum, gives a volume emissivity $j_\nu \propto
\nu^{-\alpha} \exp(-(\nu / \nu_m)^{\alpha_{\rm path}/2})$.
The function $\nu j_\nu$ has a maximum at 
\begin{equation}
\nu_{ps} = 
  \left( {3 - \sigma} \over \alpha_{\rm path} \right)^{2/\alpha_{\rm path}}
  \nu_m \equiv C_{\rm path} \nu_m.
\end{equation}
For $\alpha_{\rm path} = 1,$ $\nu_p = \nu_m,$ but if $\alpha_{\rm path} < 1,$
the increased broadening of the turnover moves the peak in 
$\nu j_\nu$ to higher frequencies, more so for flatter electron
spectra (lower $\sigma$).  (For $\sigma = 2$ and $\alpha_{\rm path} = 0.5,
C_{\rm path} = 16.$)  We can then use the observed ratio of peak
inverse-Compton frequency to peak synchrotron frequency
$R_\nu \equiv \nu_{pi} / \nu_{ps}$ to solve for the magnetic field:
\begin{equation}
B = 9 \times 10^4 C_{\rm path}^{-1} 
    \left( {\nu_{pi} \over \nu_{ps}} \right)^{-1} \ {\rm G}
\equiv 9 \times 10^4 C_{\rm path}^{-1} R_\nu^{-1} \ {\rm G}.
\end{equation}

We can write down the ratio of synchrotron to inverse Compton
emissivity at a given frequency (assuming a power law electron
spectrum and neglecting Klein-Nishina effects, as is appropriate below
the peak frequencies) from, e.g., \citet{pacholczyk70,rybicki79}, 
for the case of a thermal photon distribution at temperature $T$ 
\begin{equation}
{j_\nu(SR) \over j_\nu(IC)} =  {c_j K B^{1 + \alpha} \nu^{-\alpha}
\over K \left( m_e c^2 \right)^{\sigma - 1} \left( 2 \pi r_0^2 \right)
\left( hc \right)^{-2}  \left( kT \right)^{(\sigma + 5)/2} F(\sigma)
\left( h\nu \right)^{-\alpha}}
\end{equation}
where $K = A_e c^{\sigma + 1}$ (so $dN/dE = K E^{-\sigma}$ far below
the cutoff).  Here $r_0$ is the classical electron radius, $h$ is
Planck's constant, and $c_j = c_5(\alpha) (1.25 \times
10^{19})^\alpha$ in the notation of  \citet{pacholczyk70} ($c_5 \sim
10^{-23}$ where cgs units are used throughout).  Also, 
\begin{equation}
F(\sigma) = {\sigma^2 + 4\sigma + 11 \over \left( \sigma+3 \right)^2
   (\sigma + 5) (\sigma + 1)} 2^{\sigma + 3}  \Gamma_f\left( {{\sigma
   + 5} \over 2} \right)  \zeta\left( {{\sigma + 5} \over 2} \right)
\end{equation}
where $\Gamma_f(z)$ is the complete gamma-function and $\zeta(z)$ is
the Riemann zeta-function \citep[see, e.g.,][]{abramowitz65}. This can
be reduced to
\begin{equation}
{j_\nu(SR) \over j_\nu(IC)} = 7.92 \times 10^{-8} \  { c_j B^{1 +
 \alpha} \left( m_e c^2 \right)^{\sigma - 1} \over \left( kT
 \right)^{(\sigma + 5)/2}  F(\sigma) h^{(\sigma - 1)/2}}.
\end{equation}
Inserting $T = 2.73$ K for the cosmic microwave background, we obtain
finally 
\begin{equation}
{j_\nu(SR) \over j_\nu(IC)} \equiv \left[ C(\sigma) \right]^{-1}  B^{1
   + \alpha},
\end{equation}
defining $C(\sigma)$ for later convenience   we have $C(2.2) = 6.8
\times 10^{-14}$, $C(2.0) = 2.1 \times 10^{-14}$,  and $C(1.7) = 3.6
\times 10^{-15}$.

Now the emitting volume of synchrotron radiation may be smaller than
that of inverse Compton radiation, if the magnetic field occupies a
fraction $f_B \le 1$ of the volume.  Then the ratio of SR to IC fluxes
is smaller by a factor $f_B$.  Including this factor and inverting
equation (B5) above, we obtain
\begin{equation}
f_B = C(\sigma) {F_\nu(SR) \over F_\nu(IC)}  B^{-(1 + \alpha)}.
\end{equation}
Estimates based on similar considerations (but for $f_B \equiv 1$)
were obtained by \citet{aharonian99}.  These results reproduce the
results of detailed model fitting to within factors of  25\% for $B$
and 2 for $f_B$.



\clearpage

\begin{deluxetable}{ll}
\scriptsize
\tablecaption{Parameters of the total and polarized intensity ATCA
images  at 1.4~GHz and 2.5~GHz. }
\tablewidth{0pt}
\tablehead{
\colhead{Parameter} & \colhead{Values} \\ 
}

\startdata
$I_{1.4}$ FWHM beam  & $46\farcs 3 \times 36\farcs 2$, P.A.=$-3.8$ \\
RMS noise in $I_{1.4}$ & 0.7 \mjb \\

$Q_{1.4}, U_{1.4}$ FWHM beam & $57\farcs 1 \times 48\farcs 8$, P.A.=$-5.1$ \\
RMS noise in $Q_{1.4}, U_{1.4}$ & 0.2 \mjb \\

$I_{2.5}$ FWHM beam  & $24\farcs 7 \times 17\farcs 9$, P.A.=$-$1.6 \\
RMS noise in $I_{2.5}$ & 0.3 \mjb \\

$Q_{2.5}, U_{2.5}$ FWHM beam & $32\farcs 5 \times 27\farcs 5$, P.A.=$-$3.1 \\
RMS noise in $Q_{2.5}, U_{2.5}$ & 0.3 \mjb \\

\enddata
\label{tab-atca}
\end{deluxetable}


\clearpage

\begin{deluxetable}{lllll}
\scriptsize
\tablecaption{Results from the power law and {\tt SRCUT} models fit to
the summed {\em Chandra} spectra of  G347.3--0.5 from bright and 
 faint regions in field~1, with the 90\% confidence ranges.}
\tablewidth{0pt}
\tablehead{
\colhead{Parameter} & \colhead{Bright} & \colhead{Faint} &
\colhead{Bright} & \colhead{Faint} \\ 
\colhead{} & \colhead{{\tt POW}} & \colhead{{\tt POW}} &
\colhead{{\tt SRCUT}\tablenotemark{a}} & \colhead{{\tt SRCUT}\tablenotemark{a}}}

\startdata

$N_{\rm H}$ (\e{21}\cm{-2}) & \multicolumn{2}{c}{$7.7^{+0.2}_{-0.3} 
                               \times 10^{21}$\cm{-2}}
                            & \multicolumn{2}{c}{$7.7\times 10^{21}$\cm{-2}} \\
$\Gamma$/$\nu_{rolloff}$ (Hz)    & $2.10^{+0.04}_{-0.03}$ 
                            & $2.53^{+0.07}_{-0.07}$
                            & $6.3^{+0.8}_{-0.6} \times 10^{17}$
                            & $1.2^{+0.3}_{-0.4} \times 10^{17}$\\
$F_X$/$S_{\nu}$ (\erg/Jy)\tablenotemark{b}   
                            & $1.12^{+0.01}_{-0.03}$\ee{-10} 
                            & $2.65^{+0.07}_{-0.13}$\ee{-11}  
                            & $4.2^{+0.4}_{-0.2}$ 
                            & $2.9^{+0.6}_{-0.4}$ \\                      
$\chi^{2}_{red}$/dof & \multicolumn{2}{c}{0.99/1077} 
                     & \multicolumn{2}{c}{1.00/1078}\\

\enddata
\tablenotetext{a}{$\alpha$ frozen to 0.60 and $N_H$ frozen to the
value from the power law fit;}
\tablenotetext{b}{$F_X$ is unabsorbed X-ray flux in 0.5--10.0~keV
band; $S_{\nu}$ is radio flux at 1~GHz.}

\label{tab-joint-fit}
\end{deluxetable}


\clearpage

\begin{deluxetable}{ll}
\scriptsize
\tablecaption{Parameters used for modeling broad-band emission from
  G347.3--0.5 with synchrotron and IC mechanism. }
\tablewidth{0pt}
\tablehead{
\colhead{Model Parameters} & \colhead{Values} }
\startdata
Derived & \\
Maximum electron energy, $E_{max}$ & 5.0$_{-0.2}^{+0.1}$ TeV \\
Electron spectral index, $\sigma$  & 2.0$_{-0.04}^{+0.05}$ \\
Broadening factor, $\alpha_{path}$  & 0.5$_{-0.02}^{+0.03}$ \\
Electron normalization, $A_e$     & 5.5$_{-0.1}^{+0.1}$\ee{-32} \\
Magnetic field strength, $B$       & 15$_{-0.03}^{+0.05}$ $\mu$G \\ 
$B$ filling factor, $f_B$     & 0.010$^{-0.003}_{+0.002}$ \\
Fixed & \\
SNR volume, $V_{E}$\tablenotemark{a} & 4.9\ee{59} \cm{-3} \\
Emission filling factor, $f_{E}$  & 0.16 \\
Outer SNR radius, $R_{O}$   & 40 pc \\
Inner SNR radius, $R_{I}$   & 34 pc \\
SNR distance, $D_{pc}$ & 6.3 kpc \\
reduced $\chi^2$/ degrees of freedom & 1.55/338  \\
\enddata

\tablenotetext{a}{$V_{E} = f_{E}~\frac{4 \pi}{3} (R_{O}^{3} -
R_{I}^3 )$ taken from \citet{ellison01}}

\label{tab-model}
\end{deluxetable}


\clearpage

\clearpage

\clearpage

\clearpage

\clearpage

\clearpage

\clearpage

\clearpage

\clearpage

\clearpage

\end{document}